\documentclass[sigconf,screen]{acmart}

\usepackage[inline]{enumitem}
\usepackage{xcolor,colortbl}
\usepackage{acronym}
\usepackage{bm}
\usepackage[skip=1mm]{caption}
\usepackage{natbib}
\usepackage{hyperref}
\hypersetup{
  colorlinks   = true,    
  urlcolor     = blue,    
  linkcolor    = blue,    
  citecolor    = red      
}

\usepackage{booktabs}
\usepackage{multirow}
\usepackage{siunitx}
\makeatletter
\let\algorithmic\@undefined
\let\endalgorithmic\@undefined
\makeatother

\usepackage{algorithm}
\usepackage{algpseudocode}
\usepackage{float} 

\usepackage{booktabs,siunitx}
\usepackage{cleveref}
\usepackage{amsmath}

\definecolor{Gray}{gray}{0.85}
\definecolor{LightCyan}{rgb}{0.88,1,1}

\newcommand{\headingnodot}[1]{\vspace*{1mm}\noindent\textbf{#1}}
\newcommand{\heading}[1]{\headingnodot{#1.}}

\AtBeginDocument{%
  }

\acrodef{DDRO}{direct document relevance optimization}
\acrodef{DSI}{differentiable search indexes}
\acrodef{GenIR}{generative information retrieval}
\acrodef{LM}{language model}
\acrodef{RLRF}{reinforcement learning from relevance feedback}
\acrodef{SFT}{supervised fine-tuning}

\copyrightyear{2026}
\acmYear{2026}
\setcopyright{cc}
\setcctype{by}
\acmConference[SIGIR '26]{Proceedings of the 49th International ACM SIGIR Conference on Research and Development in Information Retrieval}{July 20--24, 2026}{Melbourne, VIC, Australia}
\acmBooktitle{Proceedings of the 49th International ACM SIGIR Conference on Research and Development in Information Retrieval (SIGIR '26), July 20--24, 2026, Melbourne, VIC, Australia}
\acmDOI{10.1145/3805712.3809725}
\acmISBN{979-8-4007-2599-9/2026/07}

\author{Kidist Amde Mekonnen}
\orcid{0009-0006-1702-9514}
\affiliation{
  \institution{University of Amsterdam}
\city{Amsterdam}
  \country{The Netherlands}
}
\email{k.a.mekonnen@uva.nl}

\author{Yubao Tang}
\orcid{0009-0003-8010-3404}
\affiliation{
  \institution{University of Amsterdam}
\city{Amsterdam}
  \country{The Netherlands}
}
\email{y.tang3@uva.nl}

\author{Maarten de Rijke}
\orcid{0000-0002-1086-0202}
\affiliation{
 \institution{University of Amsterdam}
 \city{Amsterdam}
 \country{The Netherlands}
}
\email{m.derijke@uva.nl}

\renewcommand{\shortauthors}{Kidist Amde Mekonnen, Yubao Tang, and Maarten de Rijke}

\begin{document}


\addtolength{\abovedisplayskip}{-1pt}
\addtolength{\belowdisplayskip}{-1pt} 
\addtolength{\abovedisplayshortskip}{-1pt} 
\addtolength{\belowdisplayshortskip}{-1pt} 


\title[Parametric Memory Head for Continual GenIR]{A Parametric Memory Head for Continual Generative Retrieval}

\begin{abstract}
Generative information retrieval (GenIR) consolidates retrieval into a single neural model that decodes document identifiers (docids) directly from queries. While this \emph{model-as-index} paradigm offers architectural simplicity, it is poorly suited to dynamic document collections. Unlike modular systems, where indexes are easily updated, GenIR’s knowledge is parametrically encoded in its weights; consequently, standard adaptation methods such as full and parameter-efficient fine-tuning can induce catastrophic forgetting. We show that sequential adaptation improves retrieval on newly added documents but substantially degrades performance on earlier slices, exposing a pronounced stability--plasticity trade-off.
To address this, we propose \emph{post-adaptation memory tuning} (PAMT), a memory-only stabilization stage that augments an adapted model with a modular parametric memory head (PMH). PAMT freezes the backbone and attaches a product-key memory with fixed addressing. During prefix-trie constrained decoding, decoder hidden states sparsely query PMH to produce residual corrections in hidden space; these corrections are mapped to score adjustments via the frozen output embedding matrix, computed only over trie-valid tokens. This guides docid generation while keeping routing and backbone parameters fixed. To limit cross-slice interference, PAMT updates only a fixed budget of memory values selected using decoding-time access statistics, prioritizing entries frequently activated by the current slice and rarely used in prior sessions.
Experiments on MS~MARCO and Natural Questions under sequential, disjoint corpus increments show that PAMT substantially improves retention on earlier slices with minimal impact on retrieval performance for newly added documents, while modifying only a sparse subset of memory values per session. 
\end{abstract}



\maketitle

\acresetall
\section{Introduction}
\label{sec:intro}

Generative information retrieval (GenIR) reformulates document retrieval as an autoregressive generation task, mapping natural language queries directly to unique document identifiers (docids) \citep{Metzler2021RethinkingS, Tay2022TransformerMA, recent}. By internalizing corpus knowledge in model parameters, GenIR enables end-to-end optimization, substantial index compression, and reuse of semantic priors from pretrained language models \citep{ zhou2022ultron, Zeng2024PlanningAI, Zeng2023ScalableAE,tang2023recent, mekonnen2025lightweight}. Training typically uses an encoder--decoder architecture with two complementary objectives: \emph{indexing}, which associates document content with docids, and \emph{retrieval}, which maps queries to docids. At inference time, GenIR employs constrained decoding (e.g., prefix tries) to restrict generation to valid identifiers.

\heading{Challenge: continual retrieval over dynamic corpora}
Real-world collections are dynamic: new documents arrive continuously, and retrieval systems must integrate them without compromising performance on previously indexed data \citep{mehta2023dsi}. This is particularly challenging for GenIR. Unlike modular IR pipelines with explicit index structures, GenIR entangles query–docid mappings in model parameters, so updating the model to accommodate new documents can disrupt established retrieval behavior. Joint retraining (or rehearsal) on old and new data is the most reliable way to preserve performance, but it is often prohibitively expensive at scale. In contrast, adapting only on newly arrived documents is efficient, but it typically causes catastrophic forgetting \citep{mccloskey1989catastrophic, french1999catastrophic}, leading to a stability--plasticity trade-off: gains on the new slice (\emph{plasticity}) often come at the cost of degraded retrieval on earlier content (\emph{stability}).

\heading{Existing approaches}
Prior work on continual GenIR mitigates forgetting \emph{during adaptation}, through replay or synthetic rehearsal, identifier design and index-side strategies, or parameter-efficient updates such as prompt tuning and adapters.
While effective, these methods typically couple plasticity and retention in a single objective, making performance under long sequences of continual updates sensitive to gradient trade-offs and sampling ratios (see Section~\ref{sec:related_work}).

\heading{Empirical motivation}
We analyze continual adaptation in GenIR by training on an initial corpus and sequentially incorporating disjoint document batches. We consider two representative docid designs: (i) semantic numeric identifiers such as product-quantized codes \citep{zhou2022ultron, CLEVER, ROGER, zhou-etal-2023-enhancing-generative, 10.1145/3626772.3661379}, and (ii) keyword-based identifiers such as URLs or titles \citep{Ren2023TOMEAT, li2023multiview, DeCao2020GENRE, tang2023semantic, mekonnen2025lightweight}. We find that parametric adaptation methods, including full fine-tuning and parameter-efficient updates such as LoRA \citep{hu2022lora}, reliably improve retrieval on newly added documents but substantially degrade performance on earlier slices. In contrast, when model parameters are frozen, expanding the constrained-decoding candidate set alone causes only modest retention drift, while zero-shot transfer to future-slice identifiers remains limited. These controls indicate that the dominant source of degradation is update-induced interference in the learned query--docid mapping rather than search-space growth alone. They also indicate that adapting to each new slice remains necessary for plasticity, even though it harms stability. This motivates a post-adaptation mechanism that improves retention after learning new slices, without re-optimizing the full backbone.

\heading{Our approach}
We propose \emph{post-adaptation memory tuning} (PAMT), a two-stage framework that decouples learning new docid mappings from post-adaptation decoder-side calibration. In Stage~1, the GenIR backbone is adapted to the newly arrived slice using a standard adaptation method. In Stage~2, the adapted backbone is frozen and augmented with a \emph{parametric memory head} (PMH), implemented as a product-key memory \citep{lample2019large}. During prefix-trie constrained decoding, PMH produces sparse hidden-space corrections whose effect on next-token scores is computed only over trie-valid tokens. To limit cross-slice interference, Stage~2 updates only a fixed budget of memory values selected using decoding-time access statistics, while keeping the backbone and PMH routing components fixed during the stabilization stage. This yields a value-only post-adaptation calibration step targeted at the docid-decoding interface, rather than further backbone optimization. Importantly, PAMT does not attempt to restore pre-adaptation routing; instead, it performs calibration under the routing pattern induced by Stage~1.

\heading{Contributions}
\begin{enumerate*}[label=(\roman*)]
\item We provide an empirical characterization of catastrophic forgetting in continual GenIR, quantifying the stability--plasticity trade-off across semantic numeric and keyword-based docid schemes under full and parameter-efficient adaptation.
\item We introduce \emph{post-adaptation memory tuning} (PAMT), an adapt-then-stabilize framework that improves retention after session-level adaptation through a memory-only stabilization stage, without replaying legacy supervision.
\item We introduce a decoder-side \emph{parametric memory head} (PMH) for continual GenIR that enables sparse post-adaptation value-only calibration under prefix-trie constrained decoding, allowing retention to be improved without further backbone updates or routing changes during Stage~2.
\end{enumerate*}
\vspace*{-2mm}
\section{Related Work}
\label{sec:related_work}

\textbf{Continual learning.}
Catastrophic forgetting \citep{mccloskey1989catastrophic, french1999catastrophic} remains a central challenge when training neural models on non-stationary data. Common mitigation strategies include experience replay \citep{robins1995catastrophic, rolnick2019experience, chaudhry2019continual}, parameter regularization such as EWC \citep{kirkpatrick2017overcoming}, and knowledge distillation \citep{li2017learning}. In GenIR, these strategies face additional constraints: replay requires storing queries and/or documents, raising privacy concerns \citep{liu2022continualunlearning} and scaling poorly over many slices \citep{delange2022continual, smith2023rehearsalfree, isele2018ser}; regularization becomes harder to tune as the docid space expands; and distillation introduces additional objectives and hyperparameters that may be sensitive to slice-level distribution shift.

\heading{Continual learning in GenIR}
Prior work adapts GenIR models to evolving corpora primarily through training-time updates. Replay-based approaches such as DSI++ \citep{mehta2023dsi} combine Sharpness-Aware Minimization with generative replay to mitigate forgetting. Other methods modify model capacity or interfaces: IncDSI \citep{incdsi} incrementally expands the output layer; PromptDSI \citep{huynh2024promptdsi} attaches learnable prompts; and MixLoRA-DSI \citep{huynh2025mixlora} introduces a dynamically expandable mixture of LoRA experts driven by distribution shifts. CLEVER \citep{CLEVER} combines incremental product quantization with memory-augmented learning and pseudo-query generation, while CorpusBrain++ \citep{Guo2024CorpusBrainAC} uses task-adaptive adapters with replay for continual pretraining. Complementary to backbone-adaptation methods, MDGR \citep{mdgr} emphasizes docid design to support search-space expansion under a fixed codebook, improving robustness to corpus growth without additional backbone training but potentially limiting plasticity when new query--docid mappings must be learned. In contrast to these training-time approaches, PAMT applies a separate post-adaptation stabilization stage through memory-only updates, without further modifying the backbone.

\heading{Memory-augmented language models and sparse memory updates}
Memory-layer language models \citep{berges2024memory} augment transformers with large key--value memories accessed sparsely, including product-key top-$k$ addressing, often as replacements or augmentations of feed-forward sublayers \citep{kim2020large}. This complements analyses viewing transformer feed-forward layers as unnormalized key--value memories \citep{geva-etal-2021-transformer}. Sparse Memory Finetuning \citep{lin2025continual} reduces forgetting under distribution shift by ranking memory slots with access-based TF--IDF statistics and updating only the selected value slots while freezing the rest of the model.
Our setting differs in both task and integration point. We target continual \emph{docid decoding} in GenIR under prefix-trie constraints, and attach a modular parametric memory head queried during constrained decoding rather than inserting memory layers inside the backbone. PAMT uses sparse memory updates as a post-adaptation calibration mechanism: the top-$p$ historically accessed rows are treated as protected legacy rows and excluded from updates, while a fixed budget of the remaining value rows is selected per session using current access and inverse historical access statistics. Unlike memory-layer finetuning inside transformer backbones, PAMT applies value-only calibration \emph{after} backbone adaptation, with hidden-space corrections projected through the frozen output embedding only onto trie-valid next-token scores.
\section{Preliminaries}
\label{sec:preliminaries}

\subsection{Generative information retrieval}
\label{subsec:genir}

Let $\mathcal{D}=\{d_1,\dots,d_{|\mathcal{D}|}\}$ be a document corpus and $\mathcal{Q}=\{q_1,\dots,q_{|\mathcal{Q}|}\}$ a set of queries. Each document $d\in\mathcal{D}$ is assigned an identifier sequence $I_d=(t_1,\dots,t_{|I_d|})$ whose tokens lie in a vocabulary $\mathcal{V}$. For semantic schemes (e.g., PQ codes), $\mathcal{V}$ is augmented with dedicated docid tokens; for keyword-based schemes (e.g., title+URL), identifiers are formed from the base tokenizer vocabulary. We denote the identifier set by $\mathcal{I}(\mathcal{D})=\{I_d : d\in\mathcal{D}\}$.

GenIR uses an encoder--decoder Transformer with parameters $\theta$ to model the conditional distribution over docid sequences:
\begin{equation}
P(I \mid q;\theta) \;=\; \prod_{k=1}^{|I|} P(t_k \mid t_{<k}, q;\theta).
\end{equation}
Training minimizes cross-entropy over supervision pairs $\mathcal{S}$:
\begin{equation}
\mathcal{L}(\theta) \;=\; -\sum_{(q,d)\in\mathcal{S}} \log P(I_d \mid q;\theta).
\end{equation}

At inference, GenIR retrieves by \emph{constrained docid decoding}: a prefix trie $\mathcal{T}(\mathcal{D})$ is constructed over $\mathcal{I}(\mathcal{D})$, and beam search is restricted to trie-valid tokens $A_k(\pi)\subseteq\mathcal{V}$ that extend the current prefix $\pi=t_{<k}$. This guarantees that every generated sequence corresponds to a valid document in $\mathcal{D}$. Documents are ranked by sequence log-probability (approximated via beam search), i.e.,
$\mathrm{Rel}(q,d)\triangleq \log P(I_d \mid q;\theta)$.

\vspace{-1mm}
\subsection{Continual adaptation task}
\label{subsec:continual_task}

We consider a corpus that evolves through a sequence of discrete sessions, with disjoint slices $\mathcal{D}_i \cap \mathcal{D}_j = \emptyset$ for $i\neq j$.

\begin{itemize}[leftmargin=*,nosep]
\item \emph{Session 0 ($t=0$):} The base model $\theta_0$ is optimized on the initial corpus $\mathcal{D}_0$ and query set $\mathcal{Q}_0$. A prefix trie $\mathcal{T}_0$ is constructed.
\item \emph{Subsequent sessions ($t \geq 1$):} A new document set $\mathcal{D}_t$ and corresponding query set $\mathcal{Q}_t$ are introduced. The cumulative corpus is $\mathcal{D}_{0:t} = \bigcup_{i=0}^{t} \mathcal{D}_i$, and the trie is expanded to $\mathcal{T}_{0:t} = \mathcal{T}(\mathcal{D}_{0:t})$.
\end{itemize}

At session $t$, the model must answer queries over $\mathcal{D}_{0:t}$. The challenge is to adapt to the new slice $\mathcal{D}_t$ (\emph{plasticity}) while maintaining retrieval quality on earlier slices $\mathcal{D}_{0:t-1}$ (\emph{stability}/retention) under an expanding trie and a shared identifier scheme. We distinguish two search-space protocols (Expanded vs.\ Fixed) later in §~\ref{par:search-space}.

\subsection{Document identifier schemes}
\label{subsec:docid_schemes}

We consider two representative types of docids.

\heading{Semantic product-quantized (SPQ) docids}
Each document is embedded and quantized using product quantization~\citep{jegou2010product}. The embedding is partitioned into $M$ sub-vectors, each quantized via a codebook of $K$ centroids learned on $\mathcal{D}_0$. The identifier $I_d$ is a length-$M$ sequence of centroid tokens, each corresponding to a dedicated docid token in $\mathcal{V}$. Codebooks are frozen after initial training; new documents in $\mathcal{D}_t$ are quantized using these existing centroids, inducing a shared semantic token set across slices.

\heading{Keyword-based (title+URL, TU) docids}
We construct variable-length docids from document metadata by concatenating the title with normalized URL components and tokenizing with the base vocabulary to obtain $I_d$. This yields interpretable docids that naturally share prefixes between related documents.
\section{Methodology}
\label{sec:method}
\begin{figure*}[t]
    \centering
    \includegraphics[width=\textwidth]{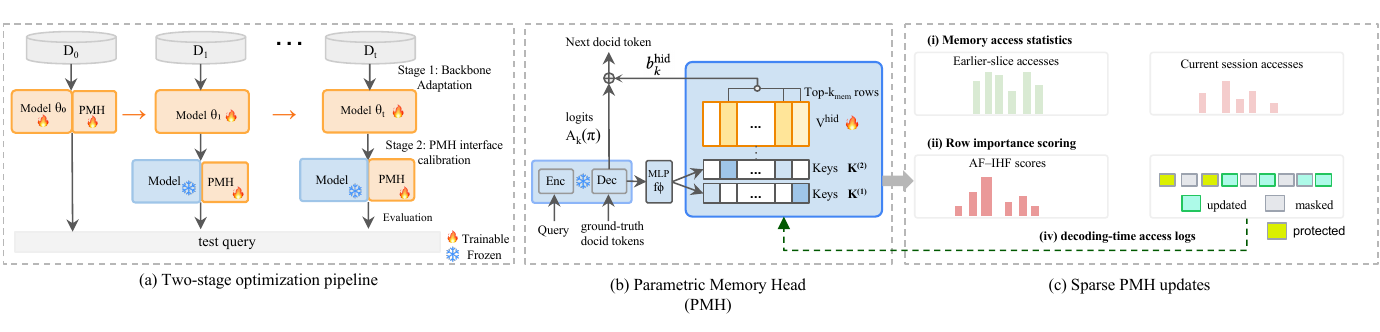}
\caption{
\textbf{Post-adaptation memory tuning (PAMT) for continual GenIR.}
\textbf{(a)} Adapt-then-stabilize pipeline: The parametric memory head (PMH) is attached and co-trained with the backbone on $\mathcal{D}_0$ to establish the initial addressing mechanism. At each session $t\!\ge\!1$, the backbone is adapted on the new slice $\mathcal{D}_t$ (Stage~1), then frozen and refined via a memory-only stabilization stage (Stage~2). Stage~2 updates a sparse subset of PMH value rows to improve retention on legacy slices $\mathcal{D}_{0:t-1}$ \emph{without gradient-based training on legacy relevance supervision}.
\textbf{(b)} Parametric memory head: During prefix-trie constrained docid decoding, decoder hidden states are projected by $f_{\phi}$ to perform product-key top-$K_{\mathrm{mem}}$ retrieval from the key tables $(K^{(1)},K^{(2)})$ and latent value table $V^{\mathrm{hid}}$. Retrieved values are aggregated into a latent correction $b^{\mathrm{hid}}_k$ in decoder hidden space; its effect on token scores is computed via the frozen output embedding matrix over trie-valid tokens $A_k(\pi)$ only.
\textbf{(c)} Sparse PMH updates: Decoding-time access logs from the current session $\mathcal{X}_t$ are contrasted with a historical access sketch to score rows using access frequency and inverse historical frequency (AF$\times$IHF), forming an update set $\Omega_t$. Sparsity is enforced by updating only $V^{\mathrm{hid}}[n]$ for $n\!\in\!\Omega_t$ while blocking gradients on protected rows $\mathcal{P}_t$; the adapted backbone, $f_{\phi}$, and keys remain fixed during Stage~2, keeping the Stage~2 routing function stable.
}
\label{fig:overview}
\end{figure*}
We propose \emph{post-adaptation memory tuning} (PAMT) for continual GenIR. As shown in Figure~\ref{fig:overview}, PAMT follows an \emph{adapt--then--stabilize} procedure that separates learning new query--docid mappings from post-adaptation decoder-side calibration. For each session $t \geq 1$, Stage~1 adapts the GenIR backbone on the arriving slice $\mathcal{D}_t$ to ensure \emph{plasticity}. Stage~2 then freezes the adapted backbone and applies a \emph{memory-only calibration} step via the parametric memory head (PMH): it updates only a fixed-budget subset of PMH value rows using current-session supervision, while excluding historically high-usage rows to preserve legacy retrieval behavior. Stage~2 does not attempt to restore pre-adaptation routing; instead, it calibrates under the routing pattern induced by Stage~1, without rehearsal on legacy slices.

\subsection{Parametric memory head (PMH)}
\label{subsec:pmh_architecture}

PMH is an external decoding-time component, co-trained with the backbone on $\mathcal{D}_0$ and later used for post-adaptation calibration of docid generation. At each decoding step, it reads the current decoder hidden state, retrieves a small set of relevant memory rows, and converts them into an additive hidden-space correction that adjusts next-token scores under prefix-trie constraints. PMH does not alter the trie constraint; it only reweights trie-valid next-token scores.

\heading{Architecture}
PMH is implemented as a parametric product-key memory (PKM) \citep{lample2019large}. Given a decoder hidden state, a learned query projection maps it into $H$ product-key query heads. These heads share a pair of product-key tables and a trainable value table:
\begin{itemize}[leftmargin=*,nosep]
\item \emph{Dual sub-key tables:} PMH stores two key tables
$K^{(1)}=\{k^{(1)}_{i}\}_{i=1}^{S}$ and
$K^{(2)}=\{k^{(2)}_{j}\}_{j=1}^{S}$, where
$k^{(1)}_{i}, k^{(2)}_{j} \in \mathbb{R}^{d_{\mathrm{key}}/2}$.
For each query head, the query is split into two halves and scored against the two shared key tables. A memory row is addressed by a compositional pair $(i,j)$, yielding a shared value address space of size $S^2$ while requiring only two $S$-way lookups per head. This product-key factorization provides large memory capacity with efficient retrieval.

\item \emph{Latent value table:} PMH has a trainable value matrix
$V^{\mathrm{hid}} \in \mathbb{R}^{S^2 \times d}$, whose rows store latent correction vectors in decoder hidden space. Each row corresponds to one flattened product-key pair $(i,j)$. We initialize
$V^{\mathrm{hid}}$ before initial co-training on $\mathcal{D}_0$; after $\mathcal{D}_0$ training, subsequent Stage~2 updates modify only selected rows of this value table. Retrieved rows are combined into a hidden-space residual whose contribution to next-token logits is computed through the output embedding matrix frozen during Stage~2.
\end{itemize}

During docid generation, the decoder produces a hidden state $h_k\in\mathbb{R}^d$ at decoding step $k$. Let $E\in\mathbb{R}^{|\mathcal{V}|\times d}$ denote the output embedding matrix of the post-adaptation backbone, which is frozen during Stage~2 and used to score next tokens.

\heading{Step 1: Product-key addressing}
At each decoding step, PMH uses $h_k$ to retrieve a small number of relevant value rows. A query module $f_\phi$ maps $h_k$ to $H$ query vectors $\{z_{k,h}\}_{h=1}^{H}$, with $z_{k,h}\in\mathbb{R}^{d_{\mathrm{key}}}$. Each query is split into two halves, $z^{(1)}_{k,h}, z^{(2)}_{k,h}\in\mathbb{R}^{d_{\mathrm{key}}/2}$, and matched against the product-key tables using dot products. A key pair $(i,j)$ is scored as
\[
u_{k,h,i,j}
=
\langle z^{(1)}_{k,h}, k^{(1)}_{i}\rangle
+
\langle z^{(2)}_{k,h}, k^{(2)}_{j}\rangle.
\]
For each head $h$, we retrieve the top-$K_{\mathrm{mem}}$ rows, where $K_{\mathrm{mem}}$ denotes the number of retrieved value rows per head:
\[
\mathcal{S}_{k,h}
=
\mathrm{TopK}\bigl(\{u_{k,h,i,j}\}_{i,j},\,K_{\mathrm{mem}}\bigr),
\qquad
\mathcal{S}_k
=
\bigcup_{h=1}^{H}\mathcal{S}_{k,h}.
\]
We flatten each key pair $(i,j)$ into a unique row index $n\in\{1,\dots,S^2\}$, for example $n=(i-1)S+j$, and compute TopK efficiently using factorized product-key lookup \citep{lample2019large}. Equivalently, $u_{k,h,n}$ denotes the score for flattened row index $n$ under head $h$.

\heading{Step 2: Latent correction construction}
For each retrieved memory row $\ell\in\mathcal{S}_{k,h}$, PMH reads a latent correction vector $V^{\mathrm{hid}}[\ell]\in\mathbb{R}^{d}$. We convert row scores into normalized weights using a softmax over the selected rows for each head:
\[
\alpha_{k,h,\ell}
=
\frac{\exp(u_{k,h,\ell})}
{\sum_{\ell'\in\mathcal{S}_{k,h}}\exp(u_{k,h,\ell'})}.
\]
PMH then forms a correction vector by taking a weighted average of the retrieved values and summing across heads:
\[
b^{\mathrm{hid}}_k
=
\sum_{h=1}^{H}
\sum_{\ell\in\mathcal{S}_{k,h}}
\alpha_{k,h,\ell}\,V^{\mathrm{hid}}[\ell],
\qquad
b^{\mathrm{hid}}_k\in\mathbb{R}^{d}.
\]
Intuitively, $b^{\mathrm{hid}}_k$ is a small additive adjustment in decoder hidden space that re-calibrates next-token scores under trie-constrained docid decoding.

\heading{Step 3: Bias-guided constrained decoding}
At decoding step $k$, PMH adjusts next-token logits only over trie-valid options. Let $\ell_k[\tau]$ denote the backbone logit for token $\tau$ at step $k$, and let $A_k(\pi)\subseteq\mathcal{V}$ be the set of trie-valid next tokens given the current prefix $\pi=t_{<k}$. For each valid token $\tau\in A_k(\pi)$, we compute
\begin{equation}
\ell'_k[\tau]
=
\ell_k[\tau]
+
\langle b^{\mathrm{hid}}_k,\,E[\tau]\rangle,
\quad \forall\, \tau \in A_k(\pi),
\label{eq:biased_logits}
\end{equation}
where $E[\tau]$ is the frozen output embedding row for token $\tau$ during Stage~2. Tokens outside $A_k(\pi)$ remain masked by the trie constraint, so PMH cannot make invalid docid continuations available. Restricting the projection to $A_k(\pi)$ keeps the additional computation proportional to the trie branching factor rather than the full vocabulary size.

\subsection{Post-adaptation memory tuning (PAMT)}
\label{subsec:method_pamt}

PAMT adds a \emph{memory-only interface calibration} stage after session-level backbone adaptation. The goal is to incorporate new document mappings while preserving index stability, \emph{without} replaying legacy relevance supervision and without further modifying the backbone.

\heading{Frozen backbone and addressing}
In Stage~2, we freeze the adapted backbone, including any session-$t$ adapters, the PMH query module $f_{\phi}$, the product-key tables $(K^{(1)},K^{(2)})$, and the output embedding matrix $E$. The only trainable component is the PMH value table $V^{\mathrm{hid}}$, from which we update a fixed budget of $m$ rows per session, corresponding to $m\cdot d$ trainable parameters. Because the correction is applied in decoder hidden space and projected to logits through the frozen output embedding, Stage~2 recalibrates the post-adaptation docid-scoring interface without further backbone updates or supervised training on legacy slices. Freezing the addressing function ensures that Stage~2 introduces no additional routing drift beyond that already induced by Stage~1 adaptation; the stage updates only the decoder-to-docid interface.

\heading{Access logs for sparse slot selection}
Let $\mathcal{X}_t$ denote the query stream, i.e., the training queries observed in session $t$. To choose which memory rows to update \emph{without} storing legacy supervision, we maintain a lightweight access log that tracks how often each memory row was retrieved in previous sessions. Concretely, we maintain a historical access-frequency counter $\mathrm{AF}^{\mathrm{hist}}_{t}(n)$ for every memory row $n\in\{1,\dots,S^2\}$. We initialize $\mathrm{AF}^{\mathrm{hist}}_{1}$ using PMH accesses collected on $\mathcal{X}_0$ after initial training on $\mathcal{D}_0$. After finishing session $t-1$, we update this counter using a \emph{binary} indicator per decoded docid sequence:
\begin{equation}
\mathrm{AF}^{\mathrm{hist}}_{t}(n)
\;\leftarrow\;
\mathrm{AF}^{\mathrm{hist}}_{t-1}(n)
\;+\;
\sum_{x \in \mathcal{X}_{t-1}}
\mathbf{1}\!\left[\exists k:\; n \in \mathcal{S}_{k}(x)\right],
\label{eq:hist_sketch_update}
\end{equation}
where $\mathcal{S}_{k}(x)$ is the set of PMH rows retrieved at step $k$ when generating the docid for query $x$. This counts how many sequences used a row at least once, rather than overweighting repeated hits within a single sequence.

At the current session $t$, we compute an analogous access count on the new slice. Current-session accesses are collected after Stage~1 adaptation, so row selection reflects the routing pattern of the adapted model used during Stage~2:
\begin{equation}
\mathrm{AF}_{t}(n)
\;=\;
\sum_{x \in \mathcal{X}_{t}}
\mathbf{1}\!\left[\exists k:\; n \in \mathcal{S}_{k}(x)\right].
\label{eq:curr_af}
\end{equation}
To make row importance comparable within the session, we normalize access frequencies as:
\begin{equation}
\widehat{\mathrm{AF}}_{t}(n)
\;=\;
\frac{\mathrm{AF}_{t}(n)}{\sum_{n'=1}^{S^2}\mathrm{AF}_{t}(n')}.
\label{eq:norm_af}
\end{equation}
We normalize current-session counts to ensure comparable row importance within the session; historical counts remain unnormalized because they serve only to downweight frequently used rows in the inverse historical frequency term. These logs store only scalar access counts, not query text or document identifiers, and do not receive gradients.

\heading{Protected set $\mathcal{P}_{t}$ (do-not-change legacy rows)}
Some memory rows are heavily used by earlier slices, so changing them is likely to hurt retention. We therefore define a \emph{protected set} $\mathcal{P}_{t}$ whose rows are frozen during Stage~2; their gradients are always zero. We select $\mathcal{P}_{t}$ as the top-$p$ fraction of rows by historical usage:
\begin{equation}
\mathcal{P}_{t}
\;=\;
\mathrm{TopP}\Bigl(\{\mathrm{AF}^{\mathrm{hist}}_{t}(n)\}_{n=1}^{S^2},\,p\Bigr),
\label{eq:protected_set}
\end{equation}
where $\mathrm{TopP}(\cdot,p)$ returns the indices of the top $p$ fraction of rows by score.

\heading{Budgeted update set $\Omega_{t}$ (update only $m$ rows)}
Stage~2 updates only a fixed budget of $m$ value rows. We choose these rows from the remaining capacity
\[
\mathcal{C}_{t}=\{1,\dots,S^2\}\setminus \mathcal{P}_{t},
\]
favoring rows that are useful for the current slice but not heavily used in the past. Following the access-based selection principle of \citet{lin2025continual}, we rank candidate rows using an AF$\times$IHF score, where AF denotes current-session access frequency and IHF denotes inverse historical frequency:
\begin{equation}
w_t(n)
\;=\;
\widehat{\mathrm{AF}}_{t}(n)\cdot
\log\frac{Z+1}{\mathrm{AF}^{\mathrm{hist}}_{t}(n)+1},
\qquad n\in\mathcal{C}_{t},
\label{eq:afihf_score}
\end{equation}
where $Z = |\mathcal{X}_{<t}|$ is the total number of queries accumulated in the historical log.\footnote{Equivalently, $Z=\sum_{s<t}|\mathcal{X}_s|$.}
The second term acts as an \emph{inverse historical frequency} (IHF),
\begin{equation}
\mathrm{IHF}_{t}(n)
=
\log\frac{Z+1}{\mathrm{AF}^{\mathrm{hist}}_{t}(n)+1},
\label{eq:ihf}
\end{equation}
which downweights rows that were frequently accessed in prior sessions. We then update only the top-$m$ rows:
\begin{equation}
\Omega_{t}
=
\mathrm{TopK}\Bigl(\{w_t(n)\}_{n\in\mathcal{C}_{t}},\,m\Bigr).
\label{eq:update_set}
\end{equation}

\heading{Stage~2 optimization and gradient masking}
Stage~2 trains \emph{only} on current-session supervision from $\mathcal{D}_{t}$, equivalently queries $\mathcal{X}_{t}$, using the biased logits from Eq.~\eqref{eq:biased_logits}. During Stage~2 training, logits are computed under teacher-forced gold prefixes. For each decoding step $k$, let $y_k$ be the gold next token, and let $\mathcal{N}_k$ be negatives sampled from the trie-valid set $A_k(\pi)\setminus\{y_k\}$. We optimize the hinge ranking objective:
\begin{equation}
\mathcal{L}_{\mathrm{rank}}
\;=\;
\sum_{k}\;\sum_{\tau^{-}\in\mathcal{N}_k}
\max\Bigl(0,\;\gamma - \ell'_k[y_k] + \ell'_k[\tau^{-}]\Bigr),
\label{eq:rank_loss}
\end{equation}
where $\gamma>0$ is a margin hyperparameter.

To enforce sparse updates, we mask gradients so that only the selected rows in $\Omega_t$ can change:
\begin{equation}
\nabla V^{\mathrm{hid}}[n]
\;\leftarrow\;
\mathbf{1}\!\left[n\in\Omega_{t}\right]\cdot \nabla V^{\mathrm{hid}}[n],
\qquad
\forall n\in\{1,\dots,S^2\}.
\label{eq:grad_mask}
\end{equation}
Rows outside $\Omega_t$ receive zero gradients. Thus, Stage~2 updates only a small, access-selected subset of memory values while keeping the backbone, addressing mechanism, and output embedding fixed.

By freezing the backbone and protecting high-usage legacy entries, PAMT concentrates session-specific corrections in less historically used memory rows, aiming to stabilize docid decoding at session $t$ while limiting interference with earlier slices, without replaying legacy relevance supervision.
\section{Experimental Setup}
\label{sec:exp}

We describe benchmarks, continual splits, model configurations, adaptation methods, and evaluation. We vary four factors, docid scheme (SPQ, TU), adaptation method (Full~FT, LoRA), search-space protocol (Expanded, Fixed), and benchmark (MS~MARCO, NQ), to analyze the plasticity--stability trade-off in Section~\ref{sec:results}.

\subsection{Datasets and continual splits}
\label{subsec:datasets}

\noindent\textbf{Benchmarks.}
We evaluate on \emph{MS~MARCO Document Ranking}~\citep{Campos2016MSMA} and \emph{Natural Questions (NQ)}~\citep{kwiatkowski-etal-2019-natural}. Following prior GenIR work~\citep{zhou2022ultron,CLEVER,mdgr,tang2023semantic,tang2024listwise,tang2024generative,tang2024bootstrapped}, we use the commonly adopted 320K-document MS~MARCO subset and NQ320K, with preprocessing from~\citet{NCI}.

\heading{Continual benchmark construction}
Each corpus is split into disjoint document slices $\mathcal{D}_0,\ldots,\mathcal{D}_n$ with $n=5$: $\mathcal{D}_0$ contains 50\% of documents, and $\mathcal{D}_1$--$\mathcal{D}_5$ each contain 10\%. At session $t$, the cumulative corpus is $\mathcal{D}_{0:t}=\bigcup_{i=0}^{t}\mathcal{D}_i$. Test queries are partitioned into $\mathcal{Q}_{\mathrm{test},0},\ldots,\mathcal{Q}_{\mathrm{test},n}$ such that each $q\in\mathcal{Q}_{\mathrm{test},t}$ has all relevant documents in $\mathcal{D}_t$ and none in $\mathcal{D}_{0:t-1}$; queries spanning multiple slices are discarded. At each session $t$, training and validation supervision is constructed only from the arriving slice $\mathcal{D}_t$. We denote the session-$t$ training queries by $\mathcal{X}_t$.

\subsection{Model architecture and document identifiers}
\label{subsec:architecture}

\noindent\textbf{GenIR backbone.}
We use \textsc{T5-base}.\footnote{\url{https://huggingface.co/google-t5/t5-base}} For SPQ docids, the decoder vocabulary is expanded from 32{,}128 to 40{,}320 tokens by adding 8{,}192 dedicated docid tokens organized into $M=32$ disjoint blocks.

\heading{Document identifier schemes}
\label{subsec:docid}
We use two docid schemes. \emph{SPQ IDs} embed each document using SentenceTransformer E5-Mistral-7B-Instruct\footnote{\url{https://huggingface.co/intfloat/e5-mistral-7b-instruct}} over the title plus the first 50 sentences, apply $\ell_2$ normalization, and PQ-code it with $M=32$ and $K=256$; the codebook is learned on $\mathcal{D}_0$ and reused for all sessions. \emph{TU IDs} concatenate title tokens (up to 20) with reversed URL path segments plus the second-level domain, tokenize with the \textsc{T5} tokenizer, and truncate to at most 100 subword tokens, with PAD/EOS removed.

\heading{Initial training on $\mathcal{D}_0$}
Models are initialized from \texttt{t5-base} and trained on $\mathcal{D}_0$ using document-to-docid pairs, pseudo-query-to-docid pairs generated via \texttt{doc2query}\footnote{\url{https://huggingface.co/castorini/doc2query-t5-base-msmarco}} with 10 pseudo-queries per document, and real query-to-docid pairs. We optimize with AdamW (lr $1{\times}10^{-3}$) for 40 epochs.

\heading{Constrained decoding}
Inference uses constrained beam search with beam size 10 over a prefix trie of valid docids. In \emph{Expanded}, decoding at session $t$ uses $\mathcal{T}(\mathcal{D}_{0:t})$ by inserting $\mathcal{I}(\mathcal{D}_t)$ into $\mathcal{T}(\mathcal{D}_{0:t-1})$. In \emph{Fixed}, decoding uses the static full-corpus trie $\mathcal{T}(\mathcal{D}_{0:n})$ at every session. Decoding is capped at $M=32$ tokens for SPQ and 100 tokens for TU.

\vspace{-1mm}
\subsection{Evaluation protocol}
\label{subsec:evaluation}

Models are first trained on $\mathcal{D}_0$ and then sequentially adapted to $\mathcal{D}_1,\ldots,\mathcal{D}_n$.

\heading{Lower-triangular evaluation}
After session $t\in\{0,\ldots,n\}$, we evaluate on all seen test query sets $\mathcal{Q}_{\mathrm{test},s}$ for $s\le t$. Let $R^{\mathrm{MRR}}_{t,s}$ denote MRR@10 on $\mathcal{Q}_{\mathrm{test},s}$ using the model state after session $t$ after PAMT when applicable. We also report $R^{\mathrm{Hit}}_{t,s}$ as Hit@10 (\%), but all aggregate continual-learning metrics are computed from $R^{\mathrm{MRR}}_{t,s}$ unless stated otherwise.

\heading{Search-space protocols}
\label{par:search-space}
We use two constrained-decoding protocols to separate model-induced forgetting from search-space expansion: \emph{Expanded} constrains decoding at session $t$ to identifiers in $\mathcal{D}_{0:t}$, while \emph{Fixed} always constrains decoding to identifiers in the full corpus $\mathcal{D}_{0:n}$. Fixed is feasible because all docids can be pre-assigned without additional learning: TU from metadata and SPQ by reusing the $\mathcal{D}_0$ codebook.

\heading{Aggregate continual-learning metrics}
\label{sec:CL_metrics}
Let $R_{t,s}\equiv R^{\mathrm{MRR}}_{t,s}$. We report:
\begin{enumerate*}[label=(\roman*)]
\item \emph{Average Performance (AP):}
$\mathrm{AP}_n=\frac{1}{n+1}\sum_{s=0}^{n}R_{n,s}$;
\item \emph{Signed Backward Transfer (BWT):}
$\mathrm{BWT}^{\pm}_n=\frac{1}{n}\sum_{s=0}^{n-1}(R_{n,s}-R_{s,s})$~\citep{lopez2017gradient};
\item \emph{Diagonal Forward Performance ($\mathrm{FWT}_{\mathrm{diag}}$):}
$\mathrm{FWT}_{\mathrm{diag},n}=\frac{1}{n}\sum_{s=1}^{n}R_{s,s}$~\citep{huynh2025mixlora,Guo2024CorpusBrainAC}, measuring average performance on each newly arrived slice immediately after adapting to that slice.
\end{enumerate*}

\heading{Stage~1 adaptation}
We consider two session-level adaptation methods. For Full~FT, $\theta_t$ denotes the backbone after adapting on $\mathcal{D}_t$, initialized from $\theta_{t-1}$; all backbone parameters are updated using target-docid negative log-likelihood with AdamW (lr $5{\times}10^{-4}$) for 3 epochs and global batch size 64. For LoRA~\citep{hu2022lora}, the backbone remains $\theta_0$ and the session-$t$ adapter is denoted $\psi_t$, so the adapted model is $(\theta_0,\psi_t)$. LoRA is applied to all linear layers using \texttt{peft}~\citep{mangrulkar2022peft} with rank $r=64$, $\alpha=128$, and dropout 0.05; adapters are initialized from $\psi_{t-1}$ and updated with AdamW (lr $1{\times}10^{-3}$), 10\% warmup, 3 epochs, and global batch size 64. During evaluation, LoRA decoding uses $(\theta_0,\psi_t)$ without merging. When applying PAMT after LoRA, both $\theta_0$ and $\psi_t$ are frozen during Stage~2.

\heading{Post-adaptation memory tuning (PAMT)}
PMH retrieves $K_{\mathrm{mem}}=32$ value rows per head to construct a latent correction $b^{\mathrm{hid}}_k\in\mathbb{R}^{d}$ (Section~\ref{subsec:pmh_architecture}); this fixed retrieval-width hyperparameter keeps PMH access sparse while allowing multiple value rows to contribute to each decoding-step correction. Its logit effect is computed via the frozen output embedding only over trie-valid tokens $A_k(\pi)$ (Eq.~\ref{eq:biased_logits}). The value table is padded to a perfect square, yielding $N_{\mathrm{mem}}=S^2$ rows for product-key addressing. Stage~2 protects the top $p=10\%$ historically accessed rows and updates a fixed budget of $m=10{,}000$ value rows per session using the AF$\times$IHF selection rule for $\Omega_t$. Stage~2 then optimizes the hinge ranking objective in Eq.~\eqref{eq:rank_loss} with margin $\gamma=0.01$, $k_{\mathrm{neg}}=8$ trie-valid negatives, and SGD (lr $1{\times}10^{-3}$) for 2 epochs with 10\% linear warmup and decay. We analyze sensitivity to the protected capacity $p$ and update budget $m$ in Section~\ref{subsec:ablation_p_m}.

\heading{Trainable parameter footprint}
Stage~1 Full~FT updates 260.16M parameters for SPQ docids and 247.58M for TU docids, while LoRA updates 25.95M adapter parameters for both schemes. Stage~2 PAMT freezes the adapted backbone and PMH addressing mechanism, updating only $m=10{,}000$ selected PMH value rows per session. With T5-base hidden size $d=768$, this gives $m\cdot d=7.68$M trainable parameters. The full PMH value table contains 160{,}000 rows on MS~MARCO and 55{,}696 rows on NQ, so the default update budget modifies 6.25\% and 17.95\% of the value table, respectively.

\heading{PMH initialization}
PMH is attached during initial training on $\mathcal{D}_0$ and trained with the standard docid negative log-likelihood loss; the hinge ranking objective is used only during Stage~2 PAMT. After session~0, the addressing mechanism, $f_\phi$ and product keys $(K^{(1)},K^{(2)})$, is frozen, and all subsequent Stage~2 updates are restricted to $V^{\mathrm{hid}}$. Thus, PAMT performs value-only calibration without introducing additional parameter-induced routing changes during Stage~2.

\heading{Baselines}
We compare against index-based retrieval and continual GenIR baselines:
\begin{enumerate*}[label=(\roman*)]
\item \emph{Lexical:} BM25~\citep{bm25} using LlamaIndex \texttt{BM25Retriever};\footnote{\url{https://developers.llamaindex.ai/python/examples/retrievers/bm25_retriever/}}
\item \emph{Dense:} DPR~\citep{Karpukhin2020DensePR} and DPR-HN~\citep{DprHN,qu-etal-2021-rocketqa} via Pyserini~\citep{Pyserini}, with the document encoder pretrained on $\mathcal{D}_0$ and frozen, and the query encoder updated per session;
\item \emph{Continual GenIR:} DSI++~\citep{mehta2023dsi}, CLEVER~\citep{CLEVER}, CorpusBrain++~\citep{Guo2024CorpusBrainAC}, PromptDSI~\citep{huynh2024promptdsi}, and MixLoRA-DSI~\citep{huynh2025mixlora}. We adapt public implementations when available; otherwise, we implement the method following the corresponding paper. For DSI++, we build on a public DSI codebase\footnote{Unofficial implementation: \url{https://github.com/ArvinZhuang/DSI-transformers}.} and add Sharpness-Aware Minimization and generative replay following~\citet{mehta2023dsi}.
\end{enumerate*}
\vspace*{-1mm}
\section{Experimental Results}
\label{sec:results}

We evaluate continual GenIR under evolving corpora through four research questions:
\begin{enumerate*}[label={(RQ\arabic*)}, font=\bfseries]
\item How do continual GenIR models trade off \emph{plasticity} and \emph{stability}, and how do adaptation method and docid design affect this trade-off?
\item To what extent does post-adaptation memory tuning (PAMT) improve \emph{stability} while preserving \emph{plasticity} on newly added documents?
\item How much degradation is driven by \emph{task-induced} effects (e.g., search-space growth and identifier transferability) versus \emph{model-induced} forgetting from parameter updates?
\item How does PAMT compare with prior continual GenIR methods under the same protocol in terms of the stability--plasticity trade-off?
\end{enumerate*}

\begin{figure}[t]
  \centering
  \includegraphics[width=\columnwidth]{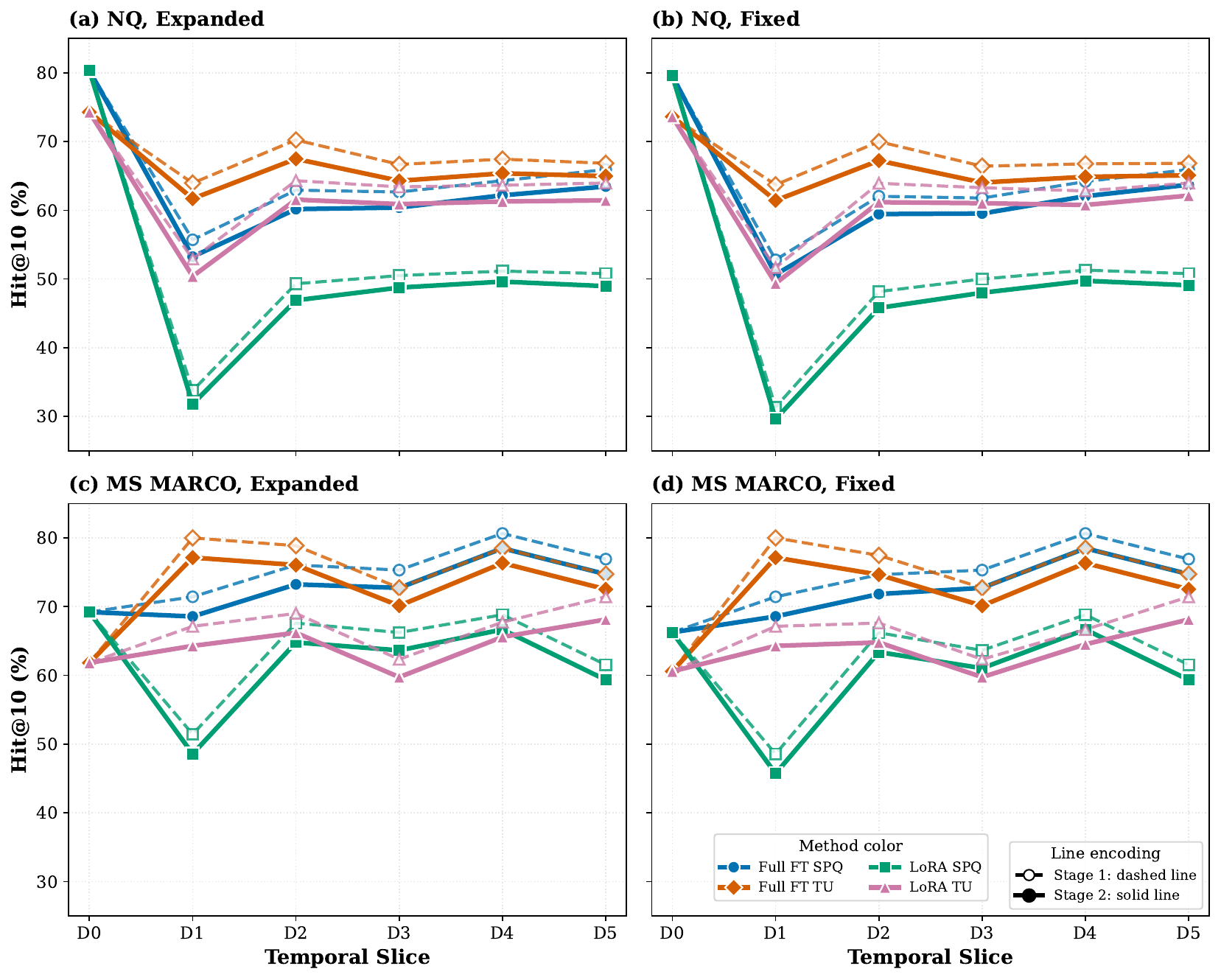}
\caption{\textbf{Stage~1 vs.\ Stage~2 across temporal slices.}
Hit@10 (\%) over slices $D_0$--$D_5$ for NQ and MS~MARCO under Expanded and Fixed protocols.
Dashed lines denote Stage~1 adaptation before PAMT; solid lines denote Stage~2 after PAMT.
The $D_0$ point corresponds to the initially trained model. Colors indicate method/docid combinations.}
  \label{fig:r1r2_lines}
  \vspace{-1mm}
\end{figure}

\begin{figure*}[t]
  \centering
  \includegraphics[width=\textwidth]{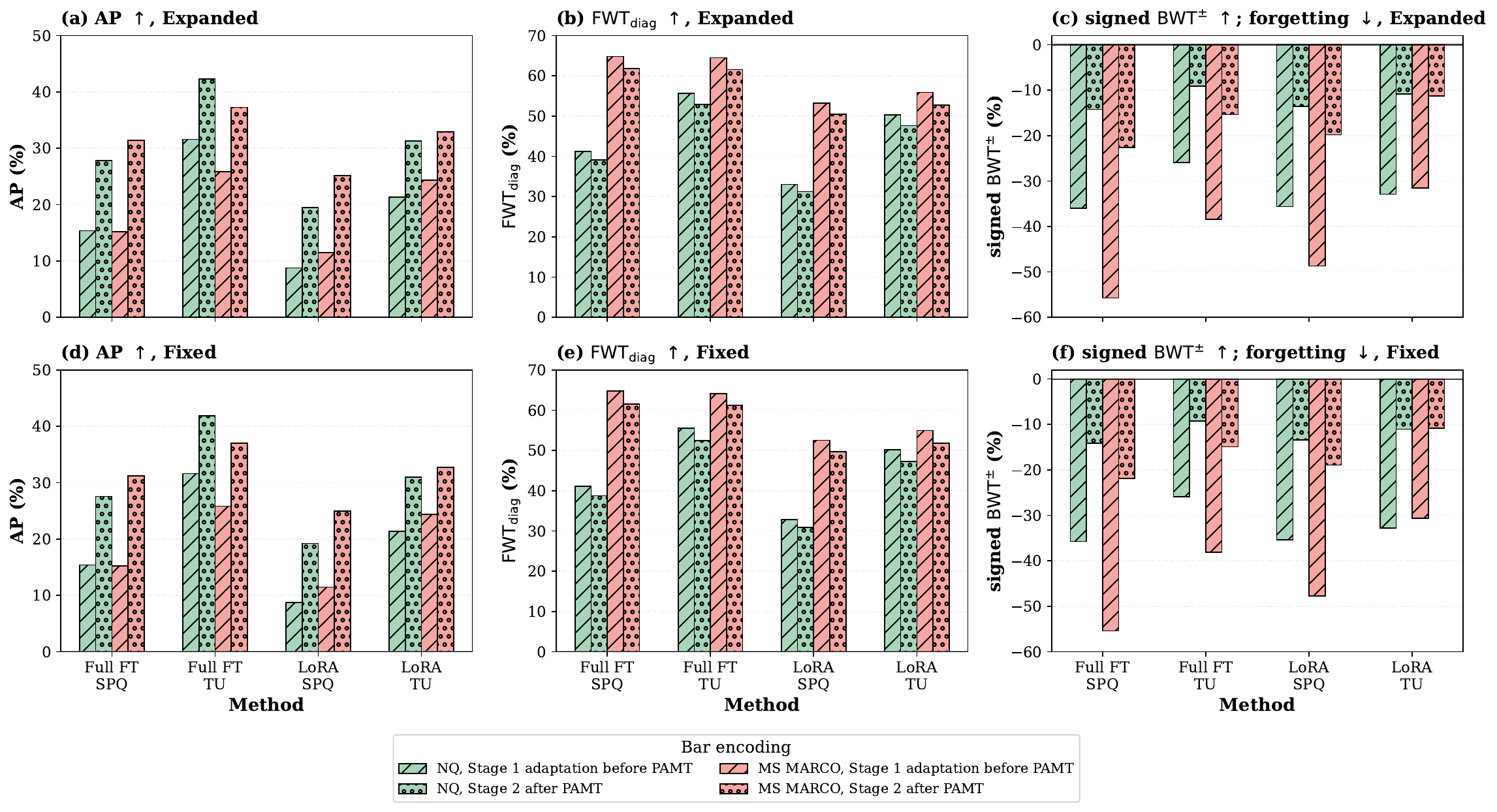}
 \caption{\textbf{Stage~1 vs.\ Stage~2 aggregate continual-learning metrics.}
AP, $\mathrm{FWT}_{\mathrm{diag}}$, and signed $\mathrm{BWT}^{\pm}$ are computed from MRR@10 and reported in percentage points for NQ and MS~MARCO under Expanded and Fixed protocols.
Hatched bars denote Stage~1 adaptation before PAMT; dotted bars denote Stage~2 after PAMT.}
  \label{fig:r1r2_bars}
  \vspace{-1mm}
\end{figure*}
\subsection{Plasticity--stability trade-off in continual GenIR}
\label{subsec:rq1_plasticity_stability}

The Stage~1 results in Figures~\ref{fig:r1r2_lines} and~\ref{fig:r1r2_bars} show a pronounced \emph{plasticity--stability trade-off} under corpus evolution (RQ1). Across both MS MARCO (MS) and Natural Questions (NQ), sequential adaptation provides plasticity on newly added slices but yields strongly negative $\mathrm{BWT}^{\pm}$ on earlier ones. Under the Expanded protocol, Full~FT with SPQ docids reaches $\mathrm{BWT}^{\pm}=-55.72$ on MS and $-35.95$ on NQ, indicating severe forgetting in continual GenIR. This pattern holds across both adaptation regimes and both docid schemes, suggesting that forgetting is a structural challenge rather than an artifact of a particular configuration.

\heading{Effect of adaptation method (Full~FT vs.\ LoRA)}
Full~FT generally provides stronger plasticity than LoRA, yielding higher diagonal Hit@10 on newly added slices and higher $\mathrm{FWT}_{\mathrm{diag}}$ under both search-space protocols. However, stronger plasticity does not consistently translate into better retention. On MS, LoRA reduces forgetting relative to Full~FT under both docid schemes (Expanded $\mathrm{BWT}^{\pm}$: SPQ $-48.70$ vs.\ $-55.72$; TU $-31.55$ vs.\ $-38.43$). On NQ, the pattern depends on the identifier scheme; for example, under Expanded with TU docids, Full~FT is more stable than LoRA ($-25.93$ vs.\ $-32.90$). These results indicate that parameter count alone does not explain forgetting; rather, the disruption of learned query--docid mappings is central.

\heading{Effect of docid design (SPQ vs.\ TU)}
Docid design strongly affects the attainable stability--plasticity trade-off. Across both datasets and adaptation methods, keyword-based TU identifiers yield less negative $\mathrm{BWT}^{\pm}$ than SPQ while maintaining competitive diagonal performance. Under Expanded with Full~FT, switching from SPQ to TU improves $\mathrm{BWT}^{\pm}$ from $-55.72$ to $-38.43$ on MS and from $-35.95$ to $-25.93$ on NQ, and substantially increases AP on NQ (15.38 $\rightarrow$ 31.59). This does not mean that TU is uniformly stronger: SPQ performs better on the initial slice $\mathcal{D}_0$, but TU yields a more favorable continual-learning profile after sequential updates. We return to this point in RQ3.

\heading{Expanded vs.\ Fixed search space}
The same qualitative pattern holds under both search-space protocols. Although Fixed can be slightly harder in earlier sessions because decoding is performed over the full identifier set from the start, method ordering and forgetting severity remain similar. For example, on MS with Full~FT--SPQ, $\mathrm{BWT}^{\pm}$ is $-55.72$ under Expanded and $-55.40$ under Fixed. Thus, candidate-set growth alone is unlikely to explain the observed degradation. The central difficulty is preserving previously learned query--docid mappings during sequential parameter updates, rather than merely decoding over a larger identifier set.

Overall, Stage~1 adaptation provides plasticity on newly added slices but fails to preserve earlier ones. This motivates a second, post-adaptation stabilization step that targets retention without further modifying the adapted backbone.

\subsection{Effectiveness of PAMT}
\label{subsec:rq2_memory_stabilization}

The Stage~2 results in Figures~\ref{fig:r1r2_lines} and~\ref{fig:r1r2_bars} show that post-adaptation memory tuning (PAMT) substantially improves \emph{stability} while preserving most of the \emph{plasticity} gained in Stage~1 (RQ2). Across both datasets, Stage~2 consistently reduces the magnitude of signed $\mathrm{BWT}^{\pm}$ relative to Stage~1. Under the Expanded protocol, Full~FT--SPQ improves from $-55.72$ to $-22.64$ on MS~MARCO and from $-35.95$ to $-14.21$ on NQ, while Full~FT--TU improves from $-38.43$ to $-15.37$ on MS~MARCO and from $-25.93$ to $-9.16$ on NQ. The same qualitative pattern holds under Fixed, indicating that the gains are not an artifact of search-space growth. This pattern is consistent across Full~FT and LoRA, and across both SPQ and TU identifiers.

\heading{Retention gains with limited impact on new-slice performance}
PAMT substantially increases AP while incurring only modest reductions in diagonal new-slice performance. Under Expanded, AP rises from 15.23 to 31.45 for Full~FT--SPQ on MS~MARCO and from 31.59 to 42.31 for Full~FT--TU on NQ, with similarly strong gains across the remaining settings. The improvements are not confined to Full~FT: under Expanded on MS~MARCO, PAMT improves $\mathrm{BWT}^{\pm}$ from $-48.70$ to $-19.82$ for LoRA--SPQ and from $-31.55$ to $-11.23$ for LoRA--TU. Thus, PAMT improves legacy retrieval while preserving most of the retrieval behavior acquired during Stage~1 adaptation. In other words, the main effect of Stage~2 is not to relearn the current slice, but to improve retention after Stage~1 while keeping the newly acquired behavior largely intact.

\heading{Interpreting the Stage~2 gains}
Stage~2 freezes the adapted backbone and PMH routing components, and updates only a sparse subset of PMH value rows. The resulting hidden-space correction is projected through the frozen output embedding and applied only to trie-valid next tokens (Eq.~\ref{eq:biased_logits}). This structurally constrains what Stage~2 can change: it can re-rank among valid continuations at decoding time, but it does not alter the backbone representations, PMH routing, or the trie constraint itself. PAMT should therefore be understood as a value-only post-adaptation calibration step on the adapted decoding interface, rather than as a mechanism for restoring pre-adaptation routing. This interpretation is consistent with the observed pattern: earlier-slice retention improves markedly, while new-slice diagonal performance is largely maintained.

\begin{table*}[t]
\centering
\caption{\textbf{RQ3 controls using the frozen $\theta_0$ checkpoint trained on $\mathcal{D}_0$.}
\emph{D0 Hit@10} is performance on $\mathcal{Q}_{\mathrm{test},0}$ with the $\mathcal{D}_0$ trie.
\emph{Frozen drop} is the absolute Hit@10 decrease on $\mathcal{Q}_{\mathrm{test},0}$ when decoding expands from $\mathcal{I}(\mathcal{D}_0)$ to $\mathcal{I}(\mathcal{D}_{0:5})$ with no parameter updates.
\emph{ZS Avg} is average strict identifier-transfer Hit@10 on $\mathcal{Q}_{\mathrm{test},t}$ for $t=1,\ldots,5$ under slice-only decoding over $\mathcal{I}(\mathcal{D}_t)$.
\emph{Coll.\ rate} is the docs-affected collision rate, i.e., the fraction of documents whose identifier is shared by at least one other document.
Candidate counts denote unique identifier sequences.}
\label{tab:rq3_controls}
\setlength{\tabcolsep}{7pt}
\renewcommand{\arraystretch}{1}
\begin{tabular}{@{}llccccc@{}}
\toprule
\textbf{Dataset} & \textbf{docid} & \textbf{D0 Hit@10} & \textbf{Frozen drop} & \textbf{Candidates} & \textbf{ZS Avg} & \textbf{Coll.\ rate} \\
\midrule
\multirow{2}{*}{MS~MARCO}
& SPQ & 69.21 & 2.95 & 152K $\rightarrow$ 301K & 5.64 & 7.45\% \\
& TU  & 61.82 & 1.23 & 158K $\rightarrow$ 315K & 14.46 & 2.37\% \\
\midrule
\multirow{2}{*}{NQ}
& SPQ & 80.34 & 0.76 & 54K $\rightarrow$ 108K & 8.97 & 3.65\% \\
& TU  & 74.28 & 0.66 & 55K $\rightarrow$ 110K & 20.25 & 1.84\% \\
\bottomrule
\end{tabular}
\end{table*}

\begin{table*}[t]
\centering
\caption{\textbf{RQ4 (Expanded): Comparison to prior approaches.}
Diagonal entries report per-slice Hit@10 (\%) after each session under the Expanded protocol, using constrained decoding over the cumulative corpus $\mathcal{D}_{0:t}$.
AP, $\mathrm{FWT}_{\mathrm{diag}}$, and signed $\mathrm{BWT}^{\pm}$ are computed from the lower-triangular MRR@10 matrix (Section~\ref{sec:CL_metrics}).
Best values in each column are underlined.}
\label{tab:expanded_msmarco_nq_hits}

\setlength{\tabcolsep}{1.8pt}
\renewcommand{\arraystretch}{1.08}

\begin{tabular}{@{}lccccccccc ccccccccc@{}}
\toprule
& \multicolumn{9}{c}{\textbf{MS~MARCO}} & \multicolumn{9}{c}{\textbf{Natural Questions}} \\
\cmidrule(lr){2-10}\cmidrule(lr){11-19}
\textbf{Method} &
D0 & D1 & D2 & D3 & D4 & D5 & AP$\uparrow$ & FWT$\uparrow$ & BWT$\uparrow$ &
D0 & D1 & D2 & D3 & D4 & D5 & AP$\uparrow$ & FWT$\uparrow$ & BWT$\uparrow$ \\
\midrule
BM25 &
67.73 & 61.43 & 70.42 & 63.64 & 56.99 & 54.95 & 30.39 & 33.56 & -5.06 &
73.78 & 71.57 & 73.52 & 68.55 & 69.06 & 67.89 & \underline{44.87} & 47.58 & -4.49 \\

DPR &
72.95 & 66.92 & 74.62 & 67.10 & 62.58 & 59.60 & 34.50 & 37.60 & -4.80 &
77.21 & 72.79 & 75.01 & 69.78 & 69.88 & 68.50 & 38.00 & 40.75 & -4.20 \\

DPR-HN &
\underline{74.35} & 68.42 & 75.82 & 68.50 & 63.98 & 61.10 & 35.50 & 38.60 & \underline{-4.60} &
78.41 & \underline{73.99} & \underline{76.21} & \underline{70.98} & \underline{71.08} & \underline{69.70} & 39.50 & 42.15 & \underline{-4.00} \\

\midrule
\multicolumn{19}{@{}l}{\emph{Non-PEFT continual GenIR baselines}} \\

DSI++ &
64.53 & 45.71 & 57.75 & 54.55 & 56.99 & 51.65 & 14.82 & 35.64 & -32.45 &
72.15 & 28.57 & 41.38 & 43.28 & 44.62 & 43.51 & 11.23 & 27.86 & -28.74 \\

CLEVER &
66.01 & 54.29 & 63.38 & 61.04 & 63.44 & 58.24 & 19.56 & 44.82 & -26.38 &
75.42 & 35.84 & 48.97 & 49.75 & 50.90 & 49.87 & 14.67 & 32.54 & -24.16 \\

\midrule
\multicolumn{19}{@{}l}{\emph{PEFT-based continual GenIR baselines}} \\

CorpusBrain++ &
63.79 & 48.57 & 59.15 & 57.14 & 59.14 & 54.95 & 17.24 & 38.96 & -29.82 &
70.58 & 31.17 & 44.14 & 45.27 & 46.15 & 45.33 & 12.84 & 29.43 & -26.58 \\

PromptDSI &
61.33 & 42.86 & 54.93 & 51.95 & 53.76 & 48.35 & 13.47 & 32.18 & -34.67 &
68.92 & 27.27 & 39.66 & 41.04 & 42.31 & 41.56 & 10.56 & 25.72 & -30.21 \\

MixLoRA-DSI &
65.27 & 61.43 & 67.61 & 64.94 & 67.74 & 64.84 & 22.83 & 51.36 & -22.94 &
77.89 & 48.57 & 58.62 & 57.96 & 59.23 & 58.31 & 24.56 & 41.87 & -18.43 \\

\midrule
\multicolumn{19}{@{}l}{\emph{Stage~1: Base adaptation (ours)}} \\

LoRA--SPQ     &
69.21 & 51.43 & 67.61 & 66.23 & 68.82 & 61.54 & 11.47 & 53.27 & -48.70 &
\underline{80.34} & 33.76 & 49.31 & 50.50 & 51.15 & 50.78 & 8.76 & 33.02  & -35.63   \\

LoRA--TU    &
61.82 & 67.14 & 69.01 & 62.34 & 67.74 & 71.43 & 24.38 & 55.88 & -31.55 &
74.28 & 52.92 & 64.31 & 63.41 & 63.64 & 63.97 & 21.36 & 50.34 & -32.90  \\

Full FT--SPQ  &
69.21 & 71.43 & 76.06 & \underline{75.32} & \underline{80.65} & \underline{76.92} & 15.23 & \underline{64.80} & -55.72 &
\underline{80.34} & 55.71 & 62.93 & 62.66 & 64.31 & 65.93 & 15.38 & 41.28 & -35.95   \\

Full FT--TU    &
61.82 & \underline{80.00} & \underline{78.87} & 72.73 & 78.49 & 74.73 & 25.83 & 64.51 & -38.43 &
74.28 & 63.96 & 70.24 & 66.67 & 67.44 & 66.84 & 31.59 & \underline{55.64} & -25.93   \\

\midrule
\multicolumn{19}{@{}l}{\emph{Stage~2: Post-Adaptation Memory Tuning (ours)}} \\

PAMT--LoRA--SPQ  &
69.21 & 48.57 & 64.79 & 63.64 & 66.67 & 59.34 & 25.18 & 50.43 & -19.82 &
\underline{80.34} & 31.82 & 46.90 & 48.75 & 49.62 & 48.96 & 19.47 & 31.24 & -13.56 \\

PAMT--LoRA--TU &
61.82 & 64.29 & 66.20 & 59.74 & 65.59 & 68.13 & 32.94 & 52.76 & -11.23 &
74.28 & 50.39 & 61.55 & 60.90 & 61.28 & 61.45 & 31.28 & 47.62 & -10.84 \\

PAMT--Full FT--SPQ  &
69.21 & 68.57 & 73.24 & 72.73 & 78.49 & 74.73 & 31.45 & 61.87 & -22.64 &
\underline{80.34} & 53.25 & 60.17 & 60.40 & 62.18 & 63.46 & 27.83 & 39.15 & -14.21 \\

PAMT--Full FT--TU &
61.82 & 77.14 & 76.06 & 70.13 & 76.34 & 72.53 & \underline{37.26} & 61.58 & -15.37 &
74.28 & 61.69 & 67.48 & 64.29 & 65.38 & 64.97 & 42.31 & 52.87 & -9.16 \\

\bottomrule
\end{tabular}
\end{table*}

\vspace*{-1mm}
\subsection{Disentangling sources of degradation}
\label{subsec:rq3_attribution}

RQ3 asks how much continual GenIR degradation is attributable to
(i) \emph{search-space growth} from increased candidate competition,
(ii) \emph{identifier transferability} to previously unseen documents, and
(iii) \emph{parameter-induced forgetting} during sequential adaptation.
To separate these factors, we fix the session-0 checkpoint $\theta_0$ trained on $\mathcal{D}_0$ and evaluate two parameter-free controls: one that expands only the constrained-decoding trie, and one that tests zero-shot transfer to future slices under slice-only decoding. Table~\ref{tab:rq3_controls} also reports docid collision rates, which help explain differences between SPQ and TU. Together, these controls separate degradation due to a changing search space, limitations of the identifier scheme, and sequential parameter updates.

\heading{Search-space growth alone causes only minor retention drift}
The first control freezes $\theta_0$ and expands only the decoding prefix trie by inserting docids from arriving slices. Retrieval is therefore performed over the cumulative identifier set $\mathcal{I}(\mathcal{D}_{0:t})$ without any parameter updates. As shown in Table~\ref{tab:rq3_controls}, retention on $\mathcal{Q}_{\mathrm{test},0}$ changes only modestly even as the searchable identifier set roughly doubles. On MS~MARCO, Hit@10 drops by 2.95\,pp for SPQ and 1.23\,pp for TU; on NQ, the drift remains below 1\,pp for both docid schemes. Thus, candidate competition alone explains only a small fraction of the retention loss observed under continual adaptation in RQ1.

\heading{Zero-shot transfer to future slices is limited and identifier-dependent}
The second control evaluates the same frozen checkpoint on future-slice queries $\mathcal{Q}_{\mathrm{test},t}$ under slice-only decoding over $\mathcal{I}(\mathcal{D}_t)$. This isolates whether a model trained only on $\mathcal{D}_0$ can decode useful identifiers for unseen documents \emph{without} adaptation. Zero-shot transfer is limited across both datasets: average strict identifier-transfer Hit@10 over $\mathcal{D}_1$--$\mathcal{D}_5$ is only 5.64 for SPQ versus 14.46 for TU on MS~MARCO, and 8.97 versus 20.25 on NQ. These results show that future-slice identifiers are not reliably decodable from $\theta_0$ alone, so learning later slices generally requires parameter updates rather than candidate-set expansion alone. They also show that identifier design already constrains continual behavior before any sequential updating occurs.

\heading{Sequential parameter updates dominate the degradation}
Compared with the two frozen controls above, continual adaptation produces substantially larger retention loss. For example, under Expanded in RQ1, Full~FT--SPQ reaches $\mathrm{BWT}^{\pm}=-55.72$ on MS~MARCO and $-35.95$ on NQ, whereas the frozen trie-expansion control produces only small Hit@10 drift. Although these quantities use different aggregate metrics, the contrast indicates that search-space growth alone is insufficient to explain the retention loss observed after sequential adaptation. The primary source of degradation is therefore training-induced interference in the learned query--docid mapping.

\heading{Why TU is more stable than SPQ under corpus evolution}
The controls in Table~\ref{tab:rq3_controls} also clarify why TU yields a more favorable continual-learning profile than SPQ. Although SPQ is stronger on the initial slice $\mathcal{D}_0$ (MS~MARCO: 69.21 vs.\ 61.82; NQ: 80.34 vs.\ 74.28), TU transfers much better to later slices. A likely reason is that TU reuses lexical subwords from the pretrained \textsc{T5} vocabulary, whereas SPQ relies on dedicated docid tokens and a fixed PQ codebook learned on $\mathcal{D}_0$. Consistent with this, SPQ exhibits substantially higher docs-affected collision rates than TU (Table~\ref{tab:rq3_controls}), which increases identifier ambiguity for later-slice documents and plausibly contributes to weaker transferability and less stable continual behavior.

Overall, RQ3 shows that continual GenIR degradation is driven primarily by update-induced interference, while docid design affects how transferable and stable the decoding space remains as the corpus evolves.

\subsection{Comparison to prior approaches}
\label{subsec:rq4_prior}

Table~\ref{tab:expanded_msmarco_nq_hits} compares PAMT with index-based retrievers and prior continual GenIR baselines under the \emph{Expanded} protocol. Index-based retrievers provide a stability reference because they maintain an explicit retrieval index rather than storing the corpus as generative docid mappings inside a single model. BM25 is non-parametric, while dense retrievers retain an explicit document index even when the query encoder is updated per session. This comparison places continual GenIR methods on a common spectrum, from retrieval architectures that largely decouple corpus growth from model-internal docid memorization to methods that must preserve legacy query--docid mappings while continuing to adapt.

\heading{Prior continual GenIR baselines still exhibit substantial forgetting}
Under the Expanded protocol, prior continual GenIR methods incur strongly negative backward transfer, indicating that updates learned for later slices disrupt legacy query--docid mappings. For example, DSI++ reaches $\mathrm{BWT}^{\pm}=-32.45$ on MS~MARCO and $-28.74$ on NQ, while CLEVER yields $-26.38$ and $-24.16$, respectively. MixLoRA-DSI is more stable than the other continual GenIR baselines, but still forgets substantially ($-22.94$ on MS~MARCO and $-18.43$ on NQ). In contrast, index-based retrievers remain much closer to $0$ (e.g., DPR-HN: $-4.60$ on MS~MARCO and $-4.00$ on NQ), reflecting the stability advantage of maintaining an explicit retrieval index rather than a model-internal generative index.

\heading{PAMT improves the continual GenIR trade-off}
Across both datasets, PAMT substantially improves stability while maintaining strong final-step effectiveness. Under Expanded, PAMT--LoRA--TU achieves $\mathrm{BWT}^{\pm}=-11.23$ on MS~MARCO and $-10.84$ on NQ, improving on all prior continual GenIR baselines in terms of stability. PAMT--Full~FT--TU attains the highest AP among continual GenIR methods, reaching 37.26 on MS~MARCO and 42.31 on NQ, while also achieving strong stability on NQ ($\mathrm{BWT}^{\pm}=-9.16$). The gains are consistent across adaptation regimes and identifier schemes. These improvements are not obtained by simply sacrificing plasticity: PAMT retains strong diagonal performance on later slices while materially improving retention on earlier ones.

The comparison also reinforces the role of identifier design. Within our GenIR variants, TU consistently yields better retention and higher AP than SPQ, while PAMT improves stability under both identifier schemes. For example, under Expanded with PAMT--Full~FT, TU improves AP from 31.45 to 37.26 on MS~MARCO and from 27.83 to 42.31 on NQ relative to SPQ. Combined with the RQ3 controls, this pattern is consistent with TU providing a more transferable and stable decoding space under corpus evolution.
Overall, PAMT materially narrows the stability gap between continual GenIR and classical retrievers through sparse post-adaptation calibration of the decoding interface, although index-based retrievers remain more stable overall.
\vspace{-4mm}
\subsection{Sensitivity to protected capacity and update budget}
\label{subsec:ablation_p_m}

We ablate the protected-row ratio $p$ and value-update budget $m$ on MS~MARCO with SPQ identifiers under the Expanded protocol as a representative sensitivity setting. The protected ratio $p$ controls how many historically accessed PMH value rows are excluded from Stage~2 updates, while $m$ controls how many remaining rows are updated. We vary $p \in \{0\%,10\%,30\%,50\%\}$ and $m \in \{2{,}000,10{,}000,50{,}000\}$; the main experiments use $p=10\%$ and $m=10{,}000$ across datasets and identifier schemes without further tuning.

\begin{table}[t]
\centering
\small
\caption{Sensitivity of PAMT to protected-row ratio $p$ and value-update budget $m$ on MS~MARCO with SPQ identifiers under the Expanded protocol. AP, $\mathrm{FWT}_{\mathrm{diag}}$, and signed $\mathrm{BWT}^{\pm}$ are computed from MRR@10; higher is better. $\dagger$ denotes the default setting.}
\label{tab:p_m_ablation}
\begin{tabular}{ccccc}
\toprule
$p$ & $m$ & AP$\uparrow$ & FWT$_{\mathrm{diag}}\uparrow$ & BWT$^{\pm}\uparrow$ \\
\midrule
0\%  & 2{,}000  & 27.80 & 59.92 & -29.84 \\
0\%  & 10{,}000 & 29.76 & 62.44 & -26.31 \\
0\%  & 50{,}000 & 29.98 & \textbf{62.73} & -25.88 \\
10\% & 2{,}000  & 29.34 & 59.46 & -25.71 \\
10\% & 10{,}000$^\dagger$ & 31.45 & 61.87 & -22.64 \\
10\% & 50{,}000 & \textbf{31.66} & 62.02 & -22.31 \\
30\% & 2{,}000  & 29.08 & 56.91 & -23.96 \\
30\% & 10{,}000 & 30.84 & 59.28 & -20.87 \\
30\% & 50{,}000 & 31.02 & 59.41 & \textbf{-20.51} \\
50\% & 2{,}000  & 26.92 & 52.43 & -24.68 \\
50\% & 10{,}000 & 28.47 & 54.96 & -21.93 \\
50\% & 50{,}000 & 28.59 & 55.11 & -21.61 \\
\bottomrule
\end{tabular}
\end{table}

Table~\ref{tab:p_m_ablation} shows a capacity--retention trade-off. Without hard protection ($p=0\%$), AF$\times$IHF selection yields worse $\mathrm{BWT}^{\pm}$ and lower AP despite marginally higher $\mathrm{FWT}_{\mathrm{diag}}$, indicating that explicit protection of historically accessed rows contributes to retention beyond AF$\times$IHF alone. Increasing $p$ improves retention up to $p=30\%$, with the best $\mathrm{BWT}^{\pm}$ obtained at $p=30\%,m=50{,}000$ ($-20.51$), but this comes at the cost of lower AP and lower $\mathrm{FWT}_{\mathrm{diag}}$ than the best AP-oriented settings. Increasing protection further to $p=50\%$ degrades all metrics relative to $p=30\%$, suggesting that over-protection leaves too little capacity for current-slice calibration. Along the budget axis, gains largely saturate beyond $m=10{,}000$: at $p=10\%$, increasing $m$ from 10{,}000 to 50{,}000 improves AP by only 0.21 and $\mathrm{BWT}^{\pm}$ by 0.33 while updating five times more rows. We therefore use $p=10\%,m=10{,}000$ as a balanced default, near the best AP while preserving update sparsity.
\vspace{-1mm}
\section{Conclusion}
\label{sec:conclusion}

We studied continual generative information retrieval (GenIR) under evolving corpora on MS~MARCO and Natural Questions using semantic product-quantized (SPQ) and title--URL (TU) identifiers. Across settings, sequential Stage~1 adaptation exhibits a clear stability--plasticity trade-off: models acquire retrieval behavior for newly added slices but incur strongly negative backward transfer on earlier ones. Additional controls show that this degradation is driven primarily by update-induced interference in the learned query--docid mapping, rather than by candidate-set growth alone.
To address this, we introduced \emph{Post-Adaptation Memory Tuning} (PAMT), an adapt-then-stabilize framework that freezes the adapted backbone and applies a memory-only stabilization stage through a modular parametric memory head. By updating only a sparse subset of memory values while keeping the backbone and PMH routing components fixed during Stage~2, PAMT improves retention without further backbone optimization or replay of legacy relevance supervision. Overall, our results indicate that a separate post-adaptation stabilization step can reduce forgetting in continual GenIR.

\heading{Limitations and future directions}
PAMT leaves several directions for future work. First,
PAMT relies on fixed design choices, notably the protected-row ratio $p$ and update budget $m$. Our ablations suggest that $m=10{,}000$ captures most Stage~2 gains, but future work could explore broader sensitivity, alternative row-selection rules, and adaptive budgets $m_t$ driven by slice novelty or PMH access divergence. Second, PAMT evaluates the retrieval-level effect of Stage~1 adaptation via signed $\mathrm{BWT}^{\pm}$, but does not quantify or constrain the induced routing changes in PMH; routing-consistency diagnostics, anchor-based routing, or access regularization could clarify when value-only calibration suffices and when routing-level intervention is needed. Third, we do not report wall-clock adaptation cost or inference latency/throughput; although PMH projections are restricted to trie-valid tokens, a full systems-level efficiency analysis remains future work. Fourth, hard protection of high-usage rows could be replaced by soft protection that scales gradients by historical importance, better accommodating stale mappings or intentional forgetting. Finally, our experiments cover 320K-scale corpora, five sessions, and two identifier schemes; longer horizons, larger corpora and backbones, broader docid designs, and joint-retraining upper bounds remain important next steps.

\begin{acks}
This research was (partially) supported by the Dutch Research Council (NWO), under project numbers 024.004.022, NWA.1389.20.183, and KICH3.LTP.20.006, and the European Union under grant agreement No. 101201510 (UNITE). Views and opinions expressed are those of the author(s) only and do not necessarily reflect those of their respective employers, funders and/or granting authorities.    
\end{acks}

\clearpage
\balance
\bibliographystyle{ACM-Reference-Format}
\bibliography{references}

@article{h,
    author = {Fu, Siyong},
    title = {A Reinforcement Learning-Based Smart Educational Environment for Higher Education},
    year = {2022},
    publisher = {IGI Global},
    address = {USA},
    volume = {19},
    number = {6},
    journal = {Int. J. e-Collab.},
    month = {dec},
    pages = {1–17},
}

@inproceedings{d,
    author = {Huang, Jin and Oosterhuis, Harrie and de Rijke, Maarten},
    booktitle = {WSDM 2022: The Fifteenth International Conference on Web Search and Data Mining},
    month = {February},
    pages = {381--289},
    publisher = {ACM},
    title = {It Is Different When Items Are Older: Debiasing Recommendations When Selection Bias and User Preferences are Dynamic},
    year = {2022}
}

@inproceedings{c,
    author = {Huang, Jin and Oosterhuis, Harrie and Cetinkaya, Bunyamin and Rood, Thijs and de Rijke, Maarten},
    booktitle = {SIGIR 2022: 45th international ACM SIGIR Conference on Research and Development in Information Retrieval},
    month = {July},
    pages = {2738--2748},
    publisher = {ACM},
    title = {State Encoders in Reinforcement Learning for Recommendation: A Reproducibility Study},
    year = {2022}
}

@inproceedings{incdsi,
  title={Incdsi: incrementally updatable document retrieval},
  author={Kishore, Varsha and Wan, Chao and Lovelace, Justin and Artzi, Yoav and Weinberger, Kilian Q},
  booktitle={International Conference on Machine Learning},
  pages={17122--17134},
  year={2023},
  organization={PMLR}
}

@inproceedings{CLEVER,
    author = {Chen, Jiangui and Zhang, Ruqing and Guo, Jiafeng and de Rijke, Maarten and Chen, Wei and Fan, Yixing and Cheng, Xueqi},
    booktitle = {CIKM 2023: 32nd ACM International Conference on Information and Knowledge Management},
    month = {October},
    pages = {306--315},
    publisher = {ACM},
    title = {Continual Learning for Generative Retrieval over Dynamic Corpora},
    year = {2023}
}

@inproceedings{mehta2023dsi,
  title={Dsi++: Updating transformer memory with new documents},
  author={Mehta, Sanket Vaibhav and Gupta, Jai and Tay, Yi and Dehghani, Mostafa and Tran, Vinh Q and Rao, Jinfeng and Najork, Marc and Strubell, Emma and Metzler, Donald},
  booktitle={Proceedings of the 2023 conference on empirical methods in natural language processing},
  pages={8198--8213},
  year={2023}
}

@inproceedings{recent,
author = {Tang, Yubao and Zhang, Ruqing and Sun, Weiwei and Guo, Jiafeng and de Rijke, Maarten},
title = {Recent Advances in Generative Information Retrieval},
year = {2024},
isbn = {9798400701726},
publisher = {Association for Computing Machinery},
address = {New York, NY, USA},
url = {https://doi.org/10.1145/3589335.3641239},
doi = {10.1145/3589335.3641239},
booktitle = {Companion Proceedings of the ACM Web Conference 2024},
pages = {1238–1241},
numpages = {4},
location = {Singapore, Singapore},
series = {WWW '24}
}

@article{Campos2016MSMA,
	author = {Daniel Fernando Campos and Tri Nguyen and Mir Rosenberg and Xia Song and Jianfeng Gao and Saurabh Tiwary and Rangan Majumder and Li Deng and Bhaskar Mitra},
	journal = {arXiv preprint arXiv:1611.09268},
	title = {{MS MARCO}: A Human Generated MAchine Reading COmprehension Dataset},
	year = {2016}
}

@article{nq,
	author = {Kwiatkowski, Tom and Palomaki, Jennimaria and Redfield, Olivia and Collins, Michael and Parikh, Ankur and Alberti, Chris and Epstein, Danielle and Polosukhin, Illia and Devlin, Jacob and Lee, Kenton and Toutanova, Kristina and Jones, Llion and Kelcey, Matthew and Chang, Ming-Wei and Dai, Andrew M. and Uszkoreit, Jakob and Le, Quoc and Petrov, Slav},
	journal = {Transactions of the Association for Computational Linguistics},
	pages = {453--466},
	publisher = {MIT Press},
	title = {Natural Questions: A Benchmark for Question Answering Research},
	volume = {7},
	year = {2019}}

@inproceedings{i,
	author = {Zhao, Xiangyu and Zhang, Liang and Ding, Zhuoye and Xia, Long and Tang, Jiliang and Yin, Dawei},
	booktitle = {KDD},
	pages = {1040--1048},
	publisher = {{ACM}},
	title = {Recommendations with Negative Feedback via Pairwise Deep Reinforcement Learning},
	year = {2018}}

@inproceedings{k,
	author = {Zheng, Guanjie and Zhang, Fuzheng and Zheng, Zihan and Xiang, Yang and Yuan, Nicholas Jing and Xie, Xing and Li, Zhenhui},
	booktitle = {WWW},
	pages = {167--176},
	publisher = {{ACM}},
	title = {DRN: A Deep Reinforcement Learning Framework for News Recommendation},
	year = {2018}}

@inproceedings{j,
	author = {Zhao, Xiangyu and Zheng, Xudong and Yang, Xiwang and Liu, Xiaobing and Tang, Jiliang},
	booktitle = {KDD},
	pages = {3319--3327},
	publisher = {{ACM}},
	title = {Jointly Learning to Recommend and Advertise},
	year = {2020}}

@inproceedings{li2023multiview,
  author    = {Li, Yongqi and Yang, Nan and Wang, Liang and Wei, Furu and Li, Wenjie},
  title     = {Multiview Identifiers Enhanced Generative Retrieval},
  booktitle = {Proceedings of the 61st Annual Meeting of the Association for Computational Linguistics (ACL 2023)},
  year      = {2023},
  pages     = {6636--6648}
}

@inproceedings{bm25,
  title={Okapi at TREC-3},
  author={Robertson, Stephen E. and Walker, Steve and Jones, Susan and Hancock-Beaulieu, Micheline and Gatford, Mike},
  booktitle={Proceedings of the Third Text REtrieval Conference (TREC-3)},
  year={1994},
  publisher={National Institute of Standards and Technology (NIST)},
  address={Gaithersburg, Maryland, USA},
  pages={109--126},
  url={https://www.microsoft.com/en-us/research/publication/okapi-at-trec-3/}
}

@inproceedings{Karpukhin2020DensePR,
    title = "Dense Passage Retrieval for Open-Domain Question Answering",
    author = "Karpukhin, Vladimir  and
      Oguz, Barlas  and
      Min, Sewon  and
      Lewis, Patrick  and
      Wu, Ledell  and
      Edunov, Sergey  and
      Chen, Danqi  and
      Yih, Wen-tau",
    editor = "Webber, Bonnie  and
      Cohn, Trevor  and
      He, Yulan  and
      Liu, Yang",
    booktitle = "Proceedings of the 2020 Conference on Empirical Methods in Natural Language Processing (EMNLP)",
    month = nov,
    year = "2020",
    address = "Online",
    publisher = "Association for Computational Linguistics",
    url = "https://aclanthology.org/2020.emnlp-main.550/",
    doi = "10.18653/v1/2020.emnlp-main.550",
    pages = "6769--6781"
}

@inproceedings{NCI,
author = {Wang, Yujing and Hou, Yingyan and Wang, Haonan and Miao, Ziming and Wu, Shibin and Sun, Hao and Chen, Qi and Xia, Yuqing and Chi, Chengmin and Zhao, Guoshuai and Liu, Zheng and Xie, Xing and Sun, Hao Allen and Deng, Weiwei and Zhang, Qi and Yang, Mao},
title = {A Neural Corpus Indexer for Document Retrieval},
year = {2022},
isbn = {9781713871088},
publisher = {Curran Associates Inc.},
address = {Red Hook, NY, USA},
booktitle = {Proceedings of the 36th International Conference on Neural Information Processing Systems},
articleno = {1856},
numpages = {15},
location = {New Orleans, LA, USA},
series = {NIPS '22}
}

@inproceedings{Tay2022TransformerMA,
author = {Tay, Yi and Tran, Vinh Q. and Dehghani, Mostafa and Ni, Jianmo and Bahri, Dara and Mehta, Harsh and Qin, Zhen and Hui, Kai and Zhao, Zhe and Gupta, Jai and Schuster, Tal and Cohen, William W. and Metzler, Donald},
title = {Transformer Memory as A Differentiable Search Index},
year = {2022},
isbn = {9781713871088},
publisher = {Curran Associates Inc.},
address = {Red Hook, NY, USA},
booktitle = {Proceedings of the 36th International Conference on Neural Information Processing Systems},
articleno = {1587},
numpages = {13},
location = {New Orleans, LA, USA},
series = {NIPS '22}
}

@inproceedings{zhou-etal-2023-enhancing-generative,
    title = "Enhancing Generative Retrieval with Reinforcement Learning from Relevance Feedback",
    author = "Zhou, Yujia  and
      Dou, Zhicheng  and
      Wen, Ji-Rong",
    editor = "Bouamor, Houda  and
      Pino, Juan  and
      Bali, Kalika",
    booktitle = "Proceedings of the 2023 Conference on Empirical Methods in Natural Language Processing",
    month = dec,
    year = "2023",
    address = "Singapore",
    publisher = "Association for Computational Linguistics",
    pages = "12481--12490",
}

@article{zhou2022ultron,
  title={Ultron: An ultimate retriever on corpus with a model-based indexer},
  author={Zhou, Yujia and Yao, Jing and Dou, Zhicheng and Wu, Ledell and Zhang, Peitian and Wen, Ji-Rong},
  journal={arXiv preprint arXiv:2208.09257},
  year={2022}
}

@inproceedings{Ren2023TOMEAT,
    title = "{TOME}: A Two-stage Approach for Model-based Retrieval",
    author = "Ren, Ruiyang  and
      Zhao, Wayne Xin  and
      Liu, Jing  and
      Wu, Hua  and
      Wen, Ji-Rong  and
      Wang, Haifeng",
    editor = "Rogers, Anna  and
      Boyd-Graber, Jordan  and
      Okazaki, Naoaki",
    booktitle = "Proceedings of the 61st Annual Meeting of the Association for Computational Linguistics (Volume 1: Long Papers)",
    month = jul,
    year = "2023",
    address = "Toronto, Canada",
    publisher = "Association for Computational Linguistics",
    url = "https://aclanthology.org/2023.acl-long.336/",
    doi = "10.18653/v1/2023.acl-long.336",
    pages = "6102--6114"
}

@inproceedings{Zeng2024PlanningAI,
author = {Zeng, Hansi and Luo, Chen and Zamani, Hamed},
title = {Planning Ahead in Generative Retrieval: Guiding Autoregressive Generation through Simultaneous Decoding},
year = {2024},
isbn = {9798400704314},
publisher = {Association for Computing Machinery},
address = {New York, NY, USA},
url = {https://doi.org/10.1145/3626772.3657746},
doi = {10.1145/3626772.3657746},
booktitle = {Proceedings of the 47th International ACM SIGIR Conference on Research and Development in Information Retrieval},
pages = {469–480},
numpages = {12},
location = {Washington DC, USA},
series = {SIGIR '24}
}

@inproceedings{Zeng2023ScalableAE,
author = {Zeng, Hansi and Luo, Chen and Jin, Bowen and Sarwar, Sheikh Muhammad and Wei, Tianxin and Zamani, Hamed},
title = {Scalable and Effective Generative Information Retrieval},
year = {2024},
isbn = {9798400701719},
publisher = {Association for Computing Machinery},
address = {New York, NY, USA},
url = {https://doi.org/10.1145/3589334.3645477},
doi = {10.1145/3589334.3645477},
booktitle = {Proceedings of the ACM Web Conference 2024},
pages = {1441–1452},
numpages = {12},
location = {Singapore, Singapore},
series = {WWW '24}
}

@article{Metzler2021RethinkingS,
author = {Metzler, Donald and Tay, Yi and Bahri, Dara and Najork, Marc},
title = {Rethinking search: making domain experts out of dilettantes},
year = {2021},
issue_date = {June 2021},
publisher = {Association for Computing Machinery},
address = {New York, NY, USA},
volume = {55},
number = {1},
issn = {0163-5840},
url = {https://doi.org/10.1145/3476415.3476428},
doi = {10.1145/3476415.3476428},
journal = {SIGIR Forum},
month = jul,
articleno = {13},
numpages = {27}
}

@article{kwiatkowski-etal-2019-natural,
    title = "Natural Questions: A Benchmark for Question Answering Research",
    author = "Kwiatkowski, Tom  and
      Palomaki, Jennimaria  and
      Redfield, Olivia  and
      Collins, Michael  and
      Parikh, Ankur  and
      Alberti, Chris  and
      Epstein, Danielle  and
      Polosukhin, Illia  and
      Devlin, Jacob  and
      Lee, Kenton  and
      Toutanova, Kristina  and
      Jones, Llion  and
      Kelcey, Matthew  and
      Chang, Ming-Wei  and
      Dai, Andrew M.  and
      Uszkoreit, Jakob  and
      Le, Quoc  and
      Petrov, Slav",
    editor = "Lee, Lillian  and
      Johnson, Mark  and
      Roark, Brian  and
      Nenkova, Ani",
    journal = "Transactions of the Association for Computational Linguistics",
    volume = "7",
    year = "2019",
    address = "Cambridge, MA",
    publisher = "MIT Press",
    url = "https://aclanthology.org/Q19-1026",
    doi = "10.1162/tacl_a_00276",
    pages = "452--466",
}

@inproceedings{DprHN,
author = {Ma, Xinyu and Guo, Jiafeng and Zhang, Ruqing and Fan, Yixing and Cheng, Xueqi},
title = {Pre-train a Discriminative Text Encoder for Dense Retrieval via Contrastive Span Prediction},
year = {2022},
isbn = {9781450387323},
publisher = {Association for Computing Machinery},
address = {New York, NY, USA},
url = {https://doi.org/10.1145/3477495.3531772},
doi = {10.1145/3477495.3531772},
booktitle = {Proceedings of the 45th International ACM SIGIR Conference on Research and Development in Information Retrieval},
pages = {848–858},
numpages = {11},
location = {Madrid, Spain},
series = {SIGIR '22}
}

@inproceedings{tang2023recent,
  title={Recent Advances in Generative Information Retrieval},
  author={Tang, Yubao and Zhang, Ruqing and Guo, Jiafeng and de Rijke, Maarten},
  booktitle={Proceedings of the Annual International ACM SIGIR Conference on Research and Development in Information Retrieval in the Asia Pacific Region},
  pages={294--297},
  year={2023}
}

@inproceedings{10.1145/3626772.3661379,
author = {Tang, Yubao and Zhang, Ruqing and Ren, Zhaochun and Guo, Jiafeng and de Rijke, Maarten},
title = {Recent Advances in Generative Information Retrieval},
year = {2024},
isbn = {9798400704314},
publisher = {Association for Computing Machinery},
address = {New York, NY, USA},
url = {https://doi.org/10.1145/3626772.3661379},
doi = {10.1145/3626772.3661379},
booktitle = {Proceedings of the 47th International ACM SIGIR Conference on Research and Development in Information Retrieval},
pages = {3005–3008},
numpages = {4},
location = {Washington DC, USA},
series = {Proceedings of the 47th International ACM SIGIR Conference on Research and Development in Information Retrieval}
}

@inproceedings{tang2024bootstrapped,
  title={Bootstrapped Pre-training with Dynamic Identifier Prediction for Generative Retrieval},
  author={Tang, Yubao and Zhang, Ruqing and Guo, Jiafeng and de Rijke, Maarten and Fan, Yixing and Cheng, Xueqi},
  booktitle={Findings of the Association for Computational Linguistics ACL 2024},
  pages={10303--10317},
  year={2024}
}

@article{tang2024listwise,
  title={Listwise Generative Retrieval Models via a Sequential Learning Process},
  author={Tang, Yubao and Zhang, Ruqing and Guo, Jiafeng and de Rijke, Maarten and Chen, Wei and Cheng, Xueqi},
  journal={ACM Transactions on Information Systems},
  volume={42},
  number={5},
  pages={1--31},
  year={2024},
  publisher={ACM New York, NY}
}

@article{ROGER,
author = {Zhou, Yujia and Yao, Jing and Dou, Zhicheng and Tu, Yiteng and Wu, Ledell and Chua, Tat-Seng and Wen, Ji-Rong},
title = {ROGER: Ranking-Oriented Generative Retrieval},
year = {2024},
issue_date = {November 2024},
publisher = {Association for Computing Machinery},
address = {New York, NY, USA},
volume = {42},
number = {6},
issn = {1046-8188},
url = {https://doi.org/10.1145/3603167},
doi = {10.1145/3603167},
journal = {ACM Trans. Inf. Syst.},
month = oct,
articleno = {155},
numpages = {25}
}

@inproceedings{mekonnen2025lightweight,
author = {Mekonnen, Kidist Amde and Tang, Yubao and de Rijke, Maarten},
title = {Lightweight and Direct Document Relevance Optimization for Generative Information Retrieval},
year = {2025},
isbn = {9798400715921},
publisher = {Association for Computing Machinery},
address = {New York, NY, USA},
url = {https://doi.org/10.1145/3726302.3730023},
doi = {10.1145/3726302.3730023},
booktitle = {Proceedings of the 48th International ACM SIGIR Conference on Research and Development in Information Retrieval},
pages = {1327–1338},
numpages = {12},
location = {Padua, Italy},
series = {SIGIR '25}
}

@article{Guo2024CorpusBrainAC,
  title={CorpusBrain++: A Continual Generative Pre-Training Framework for Knowledge-Intensive Language Tasks},
  author={Jiafeng Guo and Changjiang Zhou and Ruqing Zhang and Jiangui Chen and Maarten de Rijke and Yixing Fan and Xueqi Cheng},
  journal={ArXiv},
  year={2024},
  volume={abs/2402.16767},
  url={https://api.semanticscholar.org/CorpusID:268032853}
}

@inproceedings{mdgr,
author = {Zhang, Zhen and Ma, Xinyu and Sun, Weiwei and Ren, Pengjie and Chen, Zhumin and Wang, Shuaiqiang and Yin, Dawei and de Rijke, Maarten and Ren, Zhaochun},
title = {Replication and Exploration of Generative Retrieval over Dynamic Corpora},
year = {2025},
isbn = {9798400715921},
publisher = {Association for Computing Machinery},
address = {New York, NY, USA},
url = {https://doi.org/10.1145/3726302.3730314},
doi = {10.1145/3726302.3730314},
booktitle = {Proceedings of the 48th International ACM SIGIR Conference on Research and Development in Information Retrieval},
pages = {3325–3334},
numpages = {10},
location = {Padua, Italy},
series = {SIGIR '25}
}

@article{huynh2024promptdsi,
  title={Promptdsi: Prompt-based rehearsal-free instance-wise incremental learning for document retrieval},
  author={Huynh, Tuan-Luc and Vu, Thuy-Trang and Wang, Weiqing and Wei, Yinwei and Le, Trung and Gasevic, Dragan and Li, Yuan-Fang and Do, Thanh-Toan},
  journal={arXiv preprint arXiv:2406.12593},
  year={2024}
}

@article{huynh2025mixlora,
  title={MixLoRA-DSI: Dynamically Expandable Mixture-of-LoRA Experts for Rehearsal-Free Generative Retrieval over Dynamic Corpora},
  author={Huynh, Tuan-Luc and Vu, Thuy-Trang and Wang, Weiqing and Le, Trung and Ga{\v{s}}evi{\'c}, Dragan and Li, Yuan-Fang and Do, Thanh-Toan},
  journal={arXiv preprint arXiv:2507.09924},
  year={2025}
}

@inproceedings{DeCao2020GENRE,
  title        = {Autoregressive Entity Retrieval},
  author       = {Nicola De Cao and Gautier Izacard and Sebastian Riedel and Fabio Petroni},
  booktitle    = {Proceedings of EMNLP},
  year         = {2020}
}

@article{lin2025continual,
  title={Continual Learning via Sparse Memory Finetuning},
  author={Lin, Jessy and Zettlemoyer, Luke and Ghosh, Gargi and Yih, Wen-Tau and Markosyan, Aram and Berges, Vincent-Pierre and O{\u{g}}uz, Barlas},
  journal={arXiv preprint arXiv:2510.15103},
  year={2025}
}

@article{french1999catastrophic,
  title={Catastrophic forgetting in connectionist networks},
  author={French, Robert M},
  journal={Trends in cognitive sciences},
  volume={3},
  number={4},
  pages={128--135},
  year={1999},
  publisher={Elsevier}
}

@incollection{mccloskey1989catastrophic,
  title={Catastrophic interference in connectionist networks: The sequential learning problem},
  author={McCloskey, Michael and Cohen, Neal J},
  booktitle={Psychology of learning and motivation},
  volume={24},
  pages={109--165},
  year={1989},
  publisher={Elsevier}
}

@article{hu2022lora,
  title={Lora: Low-rank adaptation of large language models.},
  author={Hu, Edward J and Shen, Yelong and Wallis, Phillip and Allen-Zhu, Zeyuan and Li, Yuanzhi and Wang, Shean and Wang, Lu and Chen, Weizhu and others},
  journal={ICLR},
  volume={1},
  number={2},
  pages={3},
  year={2022}
}

@inbook{rolnick2019experience,
author = {Rolnick, David and Ahuja, Arun and Schwarz, Jonathan and Lillicrap, Timothy P. and Wayne, Greg},
title = {Experience replay for continual learning},
year = {2019},
publisher = {Curran Associates Inc.},
address = {Red Hook, NY, USA},
booktitle = {Proceedings of the 33rd International Conference on Neural Information Processing Systems},
articleno = {32},
numpages = {11}
}

@article{kirkpatrick2017overcoming,
  title={Overcoming catastrophic forgetting in neural networks},
  author={Kirkpatrick, James and Pascanu, Razvan and Rabinowitz, Neil and Veness, Joel and Desjardins, Guillaume and Rusu, Andrei A and Milan, Kieran and Quan, John and Ramalho, Tiago and Grabska-Barwinska, Agnieszka and others},
  journal={Proceedings of the national academy of sciences},
  volume={114},
  number={13},
  pages={3521--3526},
  year={2017},
  publisher={National Acad Sciences}
}

@inproceedings{li2017learning,
  title={Learning without forgetting},
  author={Li, Zhizhong and Hoiem, Derek},
  booktitle={IEEE transactions on pattern analysis and machine intelligence},
  volume={40},
  number={12},
  pages={2935--2947},
  year={2017},
  publisher={IEEE}
}

@inproceedings{lopez2017gradient,
  title={Gradient episodic memory for continual learning},
  author={Lopez-Paz, David and Ranzato, Marc'Aurelio},
  booktitle={Advances in Neural Information Processing Systems},
  pages={6467--6476},
  year={2017}
}

@inproceedings{qu-etal-2021-rocketqa,
    title = "{R}ocket{QA}: An Optimized Training Approach to Dense Passage Retrieval for Open-Domain Question Answering",
    author = "Qu, Yingqi  and
      Ding, Yuchen  and
      Liu, Jing  and
      Liu, Kai  and
      Ren, Ruiyang  and
      Zhao, Wayne Xin  and
      Dong, Daxiang  and
      Wu, Hua  and
      Wang, Haifeng",
    editor = "Toutanova, Kristina  and
      Rumshisky, Anna  and
      Zettlemoyer, Luke  and
      Hakkani-Tur, Dilek  and
      Beltagy, Iz  and
      Bethard, Steven  and
      Cotterell, Ryan  and
      Chakraborty, Tanmoy  and
      Zhou, Yichao",
    booktitle = "Proceedings of the 2021 Conference of the North American Chapter of the Association for Computational Linguistics: Human Language Technologies",
    month = jun,
    year = "2021",
    address = "Online",
    publisher = "Association for Computational Linguistics",
    url = "https://aclanthology.org/2021.naacl-main.466/",
    doi = "10.18653/v1/2021.naacl-main.466",
    pages = "5835--5847"
}

@misc{mangrulkar2022peft,
  title =        {{PEFT}: State-of-the-art Parameter-Efficient Fine-Tuning methods},
  author =       {Sourab Mangrulkar and Sylvain Gugger and Lysandre Debut and Younes Belkada and Sayak Paul and Benjamin Bossan and Marian Tietz},
  howpublished = {\url{https://github.com/huggingface/peft}},
  year =         {2022}
}

@article{robins1995catastrophic,
  title={Catastrophic Forgetting, Rehearsal and Pseudorehearsal},
  author={Anthony V. Robins},
  journal={Connect. Sci.},
  year={1995},
  volume={7},
  pages={123-146},
  url={https://api.semanticscholar.org/CorpusID:22882861}
}

@inproceedings{chaudhry2019continual,
  title={Continual learning with tiny episodic memories},
  author={Chaudhry, Arslan and Ranzato, Marc'Aurelio and Rohrbach, Marcus and Elhoseiny, Mohamed},
  booktitle={ICML},
  year={2019}
}

@article{berges2024memory,
  title={Memory layers at scale},
  author={Berges, Vincent-Pierre and O{\u{g}}uz, Barlas and Haziza, Daniel and Yih, Wen-tau and Zettlemoyer, Luke and Ghosh, Gargi},
  journal={arXiv preprint arXiv:2412.09764},
  year={2024}
}

@article{lample2019large,
  title={Large memory layers with product keys},
  author={Lample, Guillaume and Sablayrolles, Alexandre and Ranzato, Marc'Aurelio and Denoyer, Ludovic and J{\'e}gou, Herv{\'e}},
  journal={Advances in Neural Information Processing Systems},
  volume={32},
  year={2019}
}

@inproceedings{geva-etal-2021-transformer,
    title = "Transformer Feed-Forward Layers Are Key-Value Memories",
    author = "Geva, Mor  and
      Schuster, Roei  and
      Berant, Jonathan  and
      Levy, Omer",
    editor = "Moens, Marie-Francine  and
      Huang, Xuanjing  and
      Specia, Lucia  and
      Yih, Scott Wen-tau",
    booktitle = "Proceedings of the 2021 Conference on Empirical Methods in Natural Language Processing",
    month = nov,
    year = "2021",
    address = "Online and Punta Cana, Dominican Republic",
    publisher = "Association for Computational Linguistics",
    url = "https://aclanthology.org/2021.emnlp-main.446/",
    doi = "10.18653/v1/2021.emnlp-main.446",
    pages = "5484--5495"
}

@inproceedings{liu2022continualunlearning,
   title={Continual learning and private unlearning},
  author={Liu, Bo and Liu, Qiang and Stone, Peter},
  booktitle={Conference on Lifelong Learning Agents},
  pages={243--254},
  year={2022},
  organization={PMLR}
}

@article{delange2022continual,
  title   = {A Continual Learning Survey: Defying Forgetting in Classification Tasks},
  author  = {De Lange, Matthias and Aljundi, Rahaf and Masana, Marc and others},
  journal = {IEEE Transactions on Pattern Analysis and Machine Intelligence},
  year    = {2022}
}

@inproceedings{smith2023rehearsalfree,
  title={A closer look at rehearsal-free continual learning},
  author={Smith, James Seale and Tian, Junjiao and Halbe, Shaunak and Hsu, Yen-Chang and Kira, Zsolt},
  booktitle={Proceedings of the IEEE/CVF conference on computer vision and pattern recognition},
  pages={2410--2420},
  year={2023}
}

@inproceedings{isele2018ser,
  title     = {Selective Experience Replay for Lifelong Learning},
  author    = {Isele, David and Cosgun, Akansel},
  booktitle = {AAAI},
  year      = {2018}
}

@article{jegou2010product,
  title={Product quantization for nearest neighbor search},
  author={Jegou, Herve and Douze, Matthijs and Schmid, Cordelia},
  journal={IEEE transactions on pattern analysis and machine intelligence},
  volume={33},
  number={1},
  pages={117--128},
  year={2010},
  publisher={IEEE}
}

@inproceedings{Pyserini,
author = {Lin, Jimmy and Ma, Xueguang and Lin, Sheng-Chieh and Yang, Jheng-Hong and Pradeep, Ronak and Nogueira, Rodrigo},
title = {Pyserini: A Python Toolkit for Reproducible Information Retrieval Research with Sparse and Dense Representations},
year = {2021},
isbn = {9781450380379},
publisher = {Association for Computing Machinery},
address = {New York, NY, USA},
url = {https://doi.org/10.1145/3404835.3463238},
doi = {10.1145/3404835.3463238},
booktitle = {Proceedings of the 44th International ACM SIGIR Conference on Research and Development in Information Retrieval},
pages = {2356–2362},
numpages = {7},
location = {Virtual Event, Canada},
series = {SIGIR '21}
}

@inproceedings{kim2020large,
  title={Large product key memory for pretrained language models},
  author={Kim, Gyuwan and Jung, Tae Hwan},
  booktitle={Findings of the Association for Computational Linguistics: EMNLP 2020},
  pages={4060--4069},
  year={2020}
}

@inproceedings{tang2023semantic,
  title={Semantic-enhanced differentiable search index inspired by learning strategies},
  author={Tang, Yubao and Zhang, Ruqing and Guo, Jiafeng and Chen, Jiangui and Zhu, Zuowei and Wang, Shuaiqiang and Yin, Dawei and Cheng, Xueqi},
  booktitle={Proceedings of the 29th ACM SIGKDD Conference on Knowledge Discovery and Data Mining},
  pages={4904--4913},
  year={2023}
}

@article{tang2024generative,
  title={Generative retrieval meets multi-graded relevance},
  author={Tang, Yubao and Zhang, Ruqing and Guo, Jiafeng and de Rijke, Maarten and Chen, Wei and Cheng, Xueqi},
  journal={Advances in Neural Information Processing Systems},
  volume={37},
  pages={72790--72817},
  year={2024}
}

\end{document}